\documentclass[prd,amssymb,aps,twocolumn,epsfig]{revtex4}
\usepackage{graphicx}
\usepackage{dcolumn}   
\usepackage{amsmath,amssymb}
\usepackage{bm}
\usepackage[usenames]{color}
\definecolor{Red}{rgb}{1.0,0,0}
\definecolor{Blue}{rgb}{0,0,1.0}
\definecolor{DarkBlue}{rgb}{0.05,0.05,0.7}
\definecolor{Green}{rgb}{0,1.0,0}
\definecolor{Purple}{rgb}{0.6,0,1.0}

\def\be{\begin{equation}}       \def\ee{\end{equation}}
\def\bea{\begin{eqnarray}}      \def\eea{\end{eqnarray}}
\def\ba{\begin{array} }
\def\ea{\end{array} }
\def\bc{\begin{center}}
\def\ec{\end{center}}
\def\bnum{\begin{enumerate} }
\def\enum{\end{enumerate}}

\def\=>{\Rightarrow}
\def\>{\rightarrow}

\date{\today}

\begin{document}
\begin{text}

\title{Testing Anomalous Color-Electric Dipole Moment of the c-Quark from $\bm{\psi'\to J/\psi+\pi^++\pi^-}$ at Beijing Spectrometer}

\author{Yu-Ping Kuang$^{1}$\footnote{ypkuang@mail.tsinghua.edu.cn}, Jian-Ping Ma$^{2}$\footnote{majp@itp.ac.cn}, Otto Nachtmann$^{3}$\footnote{o.nachtmann@thphys.uni-heidelberg.de}, Wan-Peng Xie$^{1}$\footnote{xwp08@mails.tsinghua.edu.cn},  Hui-Huo Zheng$^{1,4}$\footnote{hzheng8@illinois.edu}}

\null\vspace{0.2cm}
\affiliation{$1$ Center for High Energy Physics, and Department of Physics, Tsinghua
University, Beijing, 100084, China}
\affiliation{$2$ Institute of Theoretical Physics, Academia Sinica, Beijing, 100190, China}
\affiliation{$3$ Institut f\"ur Theoretische Physik, Philosophenweg 16, 69120, Heidelberg, Germany}
\affiliation{$4$ Department of Physics, University of Illinois at Urbana-Champaign, 1110 West Green Street, Urbana, Illinois 61801-3080, USA}
\begin{abstract}

If the c quark has an anomalous color-electric dipole moment (CEDM), it may serve as a new source of CP violation.
The strength of such a CP violation depends on the size of the CEDM, $d'_c$. We propose two
effective ways of testing it from the large sample of $\psi'\to J/\psi+\pi^++\pi^-$ at the Beijing Spectrometer, and the obtained result, $\left| d'_c\right|<3\times10^{-14}$ e\,cm ($95\%$ C.L.), gives the first experimentally determined upper bound on the CEDM of the c quark.

\null\noindent{PACS numbers: 14.65.Dw, 13.20.Gd, 13.30.Eg}
\end{abstract}

\null\noindent{\null\hspace{5cm}TUHEP-TH-12173}

\maketitle

\section{Introduction}

Searching for new sources of CP violation beyond the standard model (SM)
is one of the currently interesting projects in particle physics. It concerns the
explanation of the asymmetry between matter and anti-matter in the universe.
There have been a lot of experimental studies on the CP violation in K-meson, B-meson and D-meson
decays. So far, these experimental results are consistent with the SM predictions \cite{PDG}.

There have been other possible new
CP violation sources under consideration, for example, the possible electric
dipole moments of quarks or leptons \cite{PDG}. In Ref.~\cite{Nachtmann}, the CP
violation effects in Z boson decays were studied. An effective interaction Lagrangian containing the relevant CP-violating terms was presented.
These included the electric and weak dipole moments and the color-electric dipole moment (CEDM) of the quarks. In the present paper, we are concerned with the CEDM of the c quark. We note that, to the CP-odd correlations considered in Ref.\,\cite{Nachtmann}, this CEDM does not contribute.
Ref.~\cite{MaPingZou} suggested a test via the decay $J/\psi\to\gamma\phi\phi$ based on
a naive quark model calculation. Unfortunately, there is no experimental data on the process $J/\psi\to\gamma\phi\phi$ so far. Ref.\,\cite{MaPingZou} only estimated the testing sensitivity from the statistics. In this paper, we propose a test via the hadronic transition
$\psi^\prime\to J/\psi+\pi^++\pi^-$ at the Beijing Spectrometer (BES) based on
the calculation of QCD multipole expansion \cite{Yan80,KYF,KY81,Kuang06} which has proved to be successful in many processes
\cite{Kuang06}. BES has accumulated a lot of 
$\psi'$ decays, and the branching ratio for
$\psi'\to J/\psi+\pi+\pi$ is about $50\%$, which gives a large sample for testing CEDM effect with certain precision providing the first experimental determination of the CEDM of the c quark.

The effective interaction Lagrangian including the CEDM proposed in Ref.~\cite{Nachtmann} is
\begin{eqnarray}                      
{\cal L}^{}_{CEDM}=-\frac{i}{2}d'_c\bar{\psi}_c\sigma^{\mu\nu}\gamma_5\frac{\lambda_a}{2}\psi_c G^a_{\mu\nu},
\label{cEDM}
\end{eqnarray}
where $d_c'$ is the strength of the CEDM,
$\sigma^{\mu\nu}=\displaystyle\frac{i}{2}[\gamma^\mu,\gamma^\nu]$,
$\gamma_5=i\gamma^0\gamma^1\gamma^2\gamma^3$, $\lambda_a$ is the Gell-Mann matrix for the color $SU(3)_c$ group, and
$G^a_{\mu\nu}=\partial_\mu G^a_\nu-\partial_\nu G^a_\mu-g_sf_{abc}G^b_\mu G^c_\nu$ is the field strength of
the gluon field.

${\cal L}^{}_{CEDM}$ affects the hadronic transition processes $\psi^\prime\to J/\psi+\pi^++\pi^-$
in two folds:
\begin{description}
\item{i.}  It contributes to the static potential between $c$ and $\bar c$, which causes the mixing between
CP-even and CP-odd $c\bar c$ bound states, i.e., both $\psi'$ and $J/\psi$ contain certain CP-odd
ingredients such as $\psi(^1P_1)$.
\item{ii.}  ${\cal L}^{}_{CEDM}$ contributes to the vertices in QCD multipole expansion, so that it affects the transition amplitudes.
\end{description}

In this paper, we shall calculate the above two contributions systematically.

We first treat ${\cal L}^{}_{CEDM}$ as a perturbation to calculate its contribution to the $c$-$\bar c$ static potential with which we calculate the energy shifts and CP violating state mixings.

We then calculate the theoretical prediction for the distribution
$d\Gamma(\psi'\to J/\psi+\pi^++\pi^-)/dM_{\pi\pi}$ and compare the obtained result with the BES data,
which leads to an upper bound of $d'_c$. Finally we construct a CP-odd operator ${\cal O}$ from the initial-state
and final-state momenta in $e^+e^-\to \psi'\to J/\psi+\pi^++\pi^-$, and calculate its expectation value $\langle {\cal O}\rangle$
under the amplitude of $e^+e^-\to \psi'\to J/\psi+\pi^++\pi^-$. Since the amplitude contains a CP-odd piece
proportional to $d'_c$, $\langle {\cal O}\rangle$ is proportional to $d'_c$. So measuring $\langle {\cal O}\rangle$ can provide another way of testing $d'_c$.
We suggest BESIII to do this measurement.

The CEDM interaction ${\cal L}_{CEDM}$ in (\ref{cEDM}) is a dim-5 operator in an effective Lagrangian with a scale parameter $\Lambda\sim{\rm TeV}$ beyond which the standard model (SM) should be replaced by new physics. The present study is at energies far below $\Lambda$ and also much below the electroweak symmetry breaking scale. See Ref.\,\cite{BN91} for a discussion of the effective
Lagrangian approach in such a case. In this paper, we concentrate on studying the contribution of ${\cal L}_{CEDM}$ to the hadronic transition  $\psi'\to J/\psi\,\pi\pi$. Here we would like to explain why other higher dimensional CP-odd operators, such as the CP-odd 3-gluonic CP-odd operator ${\cal O}_G=-\displaystyle\frac{C}{6}f_{abc}G^{a}_{\mu\rho}G_{\nu}^{b\;\rho} G^{c}_{\lambda\sigma}\epsilon^{\mu\nu\lambda\sigma}$ \cite{Weinberg89}, need not be included in this study.
In an effective Lagrangian theory,
an operator with dimension $4+n$ is always matched by $1/\Lambda^n$ from its coefficient. Let us first look at the dim-5 operator ${\cal L}_{CEDM}$. Comparing it with the dim-4 SM quark-gluon interaction, we see that the extra dimension of ${\cal L}_{CEDM}$ comes from the extra derivative on the gluon field, i.e., from the gluon momentum $k$. In the transition $\psi'\to J/\psi\,\pi\pi$, $k<M_{\psi'}-M_{J/\psi}=590$ MeV. Thus ${\cal L}_{CEDM}$ is suppressed by $k/\Lambda$ relative to the SM quark-gluon interaction. Next we look at the dim-6 CP-odd operator $O_G$. Comparing it with the dim-4 SM triple-gluon interaction, we see that the two extra dimensions of $O_G$ come from two extra derivatives on two gluon fields. Thus $O_G$ is suppressed by $k^2/\Lambda^2$ relative to the SM triple-gluon interaction which is of the same order as the SM quark-gluon interaction. So, $O_G$ is suppressed by $k/\Lambda< 5.9\times 10^{-4}$ relative to ${\cal L}_{CEDM}$. There have been many papers estimating the magnitude of the coefficient $C$ in $O_G$ \cite{NS00}\cite{O_G}, and showing that $C$ is really very small. Therefore, theoretically, it is reasonable to take only the leading dim-5 CP-odd operator ${\cal L}_{CEDM}$ into account, and ignore all the higher dimensional CP-odd operators such as $O_G$ in the present study.

This paper is organized as follows. In Sec.\,II, we calculate the contribution of ${\cal L}^{}_{CEDM}$ to the potential between heavy quark
and anti-quark, and treat it as a perturbation to calculate the energy shifts and state mixings caused by this contribution. We shall see that
both $J/\psi$ and $\psi'$ contain the CP-violating ingredient $\psi(1^1P_1)$, etc. These mixed quarkonium states define the initial- and final-state
in the transition $\psi^\prime\to J/\psi+\pi^++\pi^-$. Then we study the contribution of ${\cal L}^{}_{CEDM}$, as a new vertex, to the QCD
multipole expansion amplitudes, and calculate all the transition amplitudes up to $O(d_c')$ in Sec.\,III. In Sec.\,IV, we calculate the total $M_{\pi\pi}$ distribution
$d\Gamma(\psi'\to J/\psi+\pi^++\pi^-)/dM_{\pi\pi}$ and compare it with the BES measured result. This leads to an upper bound of $d'_c$ which is the strongest bound obtained so far. In Sec.\,V, we propose the alternative way of determining $d'_c$ from the experimental data on $\langle {\cal O}\rangle$. Sec.\,VI is a concluding remark.

\section{Static Potential and State Mixing}
\subsection{Derivation of the Potential}

Since $d'_c$ is supposed to be small, the ${\cal L}^{}_{CEDM}$ contributions to the potential between $c$ and $\bar c$ can be calculated
by perturbation similar to the derivation of the Coulomb potential in quantum electrodynamics \cite{Peskin}.
Let the conventional heavy quark potential be $V_0$, and the ${\cal L}^{}_{CEDM}$ contributed potential be $V_1$. The total
potential is
\bea                                            
V=V_0+V_1.
\eea
In the following, we take $V_0$ to be a QCD motivated potential, such as the Cornell potential (the simplest one) \cite{Cornell} or
the Chen-Kuang potential (more QCD, and better phenomenological predictions) \cite{CK}. Note that the short distance behavior of the Cornell potential is the hardest (steepest) among the QCD motivated potentials, while that of the CK potential is the softest (flattest). Thus comparing the results in the two potential models, we can see the model dependence of the result. Now we calculate $V_1$ to lowest order perturbation.
The Feynman diagrams for the ${\cal L}^{}_{CEDM}$ contributins to $V_1$ are showm in FIG.~\ref{FD for V_1},
where the normal vertex 
is $\displaystyle{-ig_s\gamma^\mu\frac{\lambda_a}{2}}$ for  $c$ and
$\displaystyle{ig_s\gamma^\nu\frac{\lambda_b}{2}}$ for $\bar{c}$. The shaded circle stands for the
CEDM vertex determined by ${\cal L}^{}_{CEDM}$.
\begin{center}
\begin{figure}[h]                                  
\includegraphics[width=8.4truecm,clip=true]{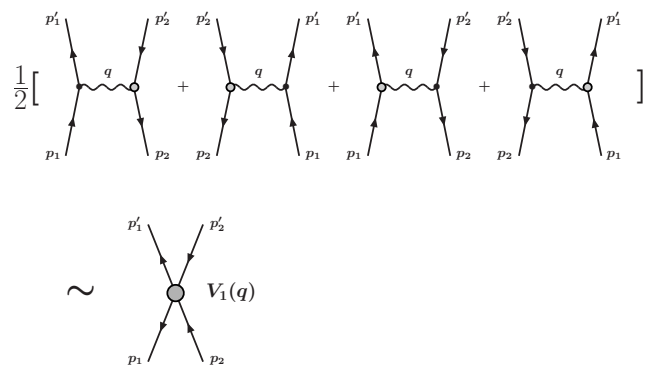}
\vspace{-0.1cm} \caption{Lowest order Feynman diagrams for $V_1$, where $\bm\bullet$ is the normal vetex and the shaded circle is the CEDM vertex.}
\label{FD for V_1}
\end{figure}
\end{center}

In the static limit, the obtained $V_1$ in the momentum representation is
\begin{eqnarray}                       
V_1(\bm q)=i\frac{4}{3}\frac{g_sd'_c}{2}
\frac{(\bm{\sigma-\bar{\sigma})\cdot\bm q}}{|\bm{q}|^2-i\epsilon}.
\label{V_1}
\end{eqnarray}
Making the Fourier transformation, we finally obtain
\bea                                  
V_1(\bm{r})&=&\frac{4g^{}_s}{3}d'_c(\bm\sigma-\bar{\bm\sigma})\cdot\bm{r}\,\delta^{(3)}_{}(\bm{r})\nonumber\\
&&-\frac{4}{3}\frac{g^{}_s}{4\pi}d'_c\frac{(\bm\sigma-\bar{\bm\sigma})\cdot\bm{r}/r}{r^2}.
\label{V_1(r)}
\end{eqnarray}
The first term serves as a repulsive core, while the second term is an attractive force.
We shall see later, especially from Eq.~(A10) in Appendix A, that {\it to first order perturbation, the first term does not make any
contributions to the energy level and the wave function corrections, so that only the second term matters}. Note that the dimension
of $d'_c$ is $m^{-1}$. So it is natrual to define
\begin{eqnarray}                 
d'_c\equiv \frac{\delta_c}{m_c},
\label{delta_c}
\end{eqnarray}
where $\delta_c$ is a dimensionless parameter. Then $V_1(\bm r)$ can be expressed by
\begin{eqnarray}                    
V_1(\bm r)&=&\frac{4g^{}_s}{3}\frac{\delta_c}{m^{}_c}(\bm\sigma-\bar{\bm\sigma})\cdot\bm{r}\,\delta^{(3)}_{}(\bm{r})\nonumber\\
&&-\frac{4}{3}\frac{g^{}_s}{4\pi}\frac{\delta_c}{m^{}_c}\frac{(\bm\sigma-\bar{\bm\sigma})\cdot\bm{r}/r}{r^2}.
\label{V_1'(r)}
\end{eqnarray}

\subsection{Energy Shift and State Mixing}

We see that $V_1(\bm r)$ contains a factor $(\bm\sigma-\bar{\bm\sigma})\cdot\bm{r}$ which flips
the quarkonium spins by $\Delta s=\pm 1$ and the quarkonium orbital angular momentum by $\Delta l=\pm 1$.
This does not change the charge conjugation but changes the parity, i.e., it violates CP. Take the
$^3S_1$ quarkonium as an example. $V_1(\bm r)$ changes this state to $^1P_1$. Thus {\it when the
potential contains $V_1(\bm r)$, the eigen-state is a mixture of the $^3S_1$ and the $^1P_1$ states}.
This affects the decays of the heavy quarkonia.

Since $\delta_c$ is supposed to be small, we can take $V_1(\bm r)$ as a perturbation. Let $E^0_{nl}$ and
$|n~^{(2s+1)}L_J\rangle_0~(s=0,1)$ be the energy eigenvalue and the wave function of the quarkonium eigenstate
with only $V_0(r)$.

To first order of $\delta_c$, the coreection to the energy eigenvalue is
\begin{eqnarray}                        
E_{nl}=E^0_{nl}
+~_0\langle n~^{(2s+1)}L_J|V_1|n~^{(2s+1)}L_J\rangle_0.
\label{E}
\end{eqnarray}
We know that $|n~^{(2s+1)}L_J\rangle_0$ is a CP eigenstate, and $V_1$ violates CP. So that the
diagonal matrix element in Eq. (\ref{E}) vanishes. Therefore {\it there is no energy shift to $O(\delta_c)$}.
Energy shift is of $O(\delta_c^2)$.

The first order wave function correction is
\begin{widetext}
\begin{eqnarray}                             
&&|n~^{(2s+1)}L_J\rangle=|n~^{(2s+1)}L_J\rangle_0
+\displaystyle\sum_{n'}\frac{_0\langle n'~^{(2(s\pm 1)+1)}(L\pm 1)_J|V_1|n~^{(2s+1)}L_J\rangle_0}
{E^0_{nl}-E^0_{n',~l\pm 1}}
\times |n'~^{(2(s\pm 1)+1)}(L\pm 1)_J\rangle_0.
\end{eqnarray}
For example,


\begin{eqnarray}                          
&&|1^3S_1\rangle=|1^3S_1\rangle_0
-\frac{_0\langle 1^1P_1|V_1|1^3S_1\rangle_0}{E^0_{1^1P_1}-E^0_{1^3S_1}}
~|1^1P_1\rangle_0+\cdots\nonumber
\\
&&|2^3S_1\rangle=|2^3S_1\rangle_0+\frac{_0\langle 1^1P_1|V_1|2^3S_1\rangle_0}{E^0_{2^3S_1}-E^0_{1^1P_1}}
~|1^1P_1\rangle_0+\cdots\nonumber
\\
\label{mixing}
&&|1^1P_1\rangle=|1^1P_1\rangle_0+\frac{_0\langle 1^3S_1|V_1|1^1P_1\rangle_0}{E^0_{1^1P_1}-E^0_{1^3S_1}}
~|1^3S_1\rangle_0
-\frac{_0\langle 2^3S_1|V_1|1^1P_1\rangle_0}{E^0_{2^3S_1}-E^0_{1^1P_1}}~|2^3S_1\rangle_0
-\frac{_0\langle 1^3D_1|V_1|1^1P_1\rangle_0}{E^0_{1^3D_1}-E^0_{1^1P_1}}
~|1^3D_1\rangle_0+\cdots\nonumber\\
&&|1^3D_1\rangle=|1^3D_1\rangle_0+\frac{_0\langle 1^1P_1|V_1|1^3D_1\rangle_0}{E^0_{1^3D_1}
-E^0_{1^1P_1}}
\times|1^1P_1\rangle_0+\cdots.
\label{mixing}
\end{eqnarray}

Here we see explicitly the mixing of the $^3S_1$ and $^1P_1$ states.

\null~~~The above expressions for the mixed states are not normalized yet. The normalized states are

\begin{eqnarray}                            
&&|1^3S_1\rangle=C_{10}^{10}~|1^3S_1\rangle_0+C_{10}^{11}~|1^1P_1\rangle_0+\cdots\nonumber\\
&&|2^3S_1\rangle=C_{20}^{20}~|2^3S_1\rangle_0+C_{20}^{11}~|1^1P_1\rangle_0+\cdots\nonumber\\
&&|1^1P_1\rangle=C_{11}^{11}~|1^1P_1\rangle_0
+C_{11}^{10}~|1^3S_1\rangle_0+C_{11}^{20}~
|2^3S_1\rangle_0
+C_{11}^{12}~|1^3D_1\rangle_0+\cdots\nonumber\\
&&|1^3D_1\rangle=C_{12}^{12}~|1^3D_1\rangle_0+C_{12}^{11}~|1^1P_1\rangle_0+\cdots,
\label{nmixing}
\end{eqnarray}
where
\begin{eqnarray}                       
&&\hspace{-0.8cm}\displaystyle C_{10}^{10}=\frac{1}{\sqrt{1+\displaystyle\left(\frac{_0\langle 1^1P_1|V_1
|1^3S_1\rangle_0}
{E^0_{1^1P_1}-E^0_{1^3S_1}}\right)^2}},
~~~~~~~~~~~~~~~~
\displaystyle C_{10}^{11}=-\frac{\displaystyle\frac{_0\langle 1^1P_1|V_1|1^3S_1\rangle_0}
{E^0_{1^1P_1}-E^0_{1^3S_1}}}{\sqrt{1+\displaystyle\left(\frac{_0\langle 1^1P_1|V_1|1^3S_1\rangle_0}
{E^0_{1^1P_1}-E^0_{1^3S_1}}\right)^2}},\nonumber\\
&&\hspace{-0.8cm}\displaystyle C_{20}^{20}=\frac{1}{\sqrt{1+\displaystyle\left(\frac{_0\langle 1^1P_1|V_1
|2^3S_1\rangle_0}
{E^0_{2^3S_1}-E^0_{1^1P_1}}\right)^2}},
~~~~~~~~~~~~~~~~
\displaystyle C_{20}^{11}=\frac{\displaystyle\frac{_0\langle 1^1P_1|V_1|2^3S_1\rangle_0}
{E^0_{2^3S_1}-E^0_{1^1P_1}}}{\sqrt{1+\displaystyle\left(\frac{_0\langle 1^1P_1|V_1|2^3S_1\rangle_0}
{E^0_{2^3S_1}-E^0_{1^1P_1}}\right)^2}},\nonumber\\
\label{mixing}
&&\hspace{-0.8cm}\displaystyle C_{11}^{11}=\frac{1}{\sqrt{1+\displaystyle\left(\frac{_0\langle 1^1P_1|V_1
|1^3S_1\rangle_0}{E^0_{1^3S_1}-E^0_{1^1P_1}}\right)^2+\displaystyle\left(\frac{_0\langle 1^1P_1|V_1
|2^3S_1\rangle_0}{E^0_{2^3S_1}-E^0_{1^1P_1}}\right)^2+
\displaystyle\left(\frac{_0\langle 1^1P_1|V_1
|1^3D_1\rangle_0}{E^0_{1^3D_1}-E^0_{1^1P_1}}\right)^2}},\nonumber
\end{eqnarray}
\end{widetext}
\begin{widetext}
\begin{eqnarray}
&&\hspace{-0.8cm}\displaystyle C_{11}^{10}=\frac{\displaystyle\frac{_0\langle 1^3S_1|V_1|1^1P_1\rangle_0}
{E^0_{1^1P_1}-E^0_{1^3S_1}}}{\sqrt{1+\displaystyle\left(\frac{_0\langle 1^1P_1|V_1
|1^3S_1\rangle_0}{E^0_{1^3S_1}-E^0_{1^1P_1}}\right)^2+\displaystyle\left(\frac{_0\langle 1^1P_1|V_1
|2^3S_1\rangle_0}{E^0_{2^3S_1}-E^0_{1^1P_1}}\right)^2+
\displaystyle\left(\frac{_0\langle 1^1P_1|V_1
|1^3D_1\rangle_0}{E^0_{1^3D_1}-E^0_{1^1P_1}}\right)^2}},\nonumber\\
&&\hspace{-0.8cm}\displaystyle C_{11}^{20}=-\frac{\displaystyle\frac{_0\langle 2^3S_1|V_1|1^1P_1\rangle_0}
{E^0_{2^3S_1}-E^0_{1^1P_1}}}{\sqrt{1+\displaystyle\left(\frac{_0\langle 1^1P_1|V_1
|1^3S_1\rangle_0}{E^0_{1^3S_1}-E^0_{1^1P_1}}\right)^2+\displaystyle\left(\frac{_0\langle 1^1P_1|V_1
|2^3S_1\rangle_0}{E^0_{2^3S_1}-E^0_{1^1P_1}}\right)^2+
\displaystyle\left(\frac{_0\langle 1^1P_1|V_1
|1^3D_1\rangle_0}{E^0_{1^3D_1}-E^0_{1^1P_1}}\right)^2}},\nonumber\\
&&
\displaystyle C_{11}^{12}=-\frac{\displaystyle\frac{_0\langle 1^3D_1|V_1|1^1P_1\rangle_0}
{E^0_{1^3D_1}-E^0_{1^1P_1}}}{\sqrt{1+\displaystyle\left(\frac{_0\langle 1^1P_1|V_1
|1^3S_1\rangle_0}{E^0_{1^3S_1}-E^0_{1^1P_1}}\right)^2+\displaystyle\left(\frac{_0\langle 1^1P_1|V_1
|2^3S_1\rangle_0}{E^0_{2^3S_1}-E^0_{1^1P_1}}\right)^2+
\displaystyle\left(\frac{_0\langle 1^1P_1|V_1
|1^3D_1\rangle_0}{E^0_{1^3D_1}-E^0_{1^1P_1}}\right)^2}}\nonumber\\
&&\hspace{-0.8cm}\displaystyle C_{12}^{12}
=\frac{1}{\sqrt{1+\displaystyle\left(\frac{_0\langle 1^1P_1|V_1|1^3D_1\rangle_0}
{E^0_{1^3D_1}-E^0_{1^1P_1}}\right)^2}},~~~~~~~~
\displaystyle C_{12}^{11}=\frac{\displaystyle\frac{_0\langle 1^1P_1|V_1|1^3D_1\rangle_0}
{E^0_{1^3D_1}-E^0_{1^1P_1}}}{\sqrt{1+\displaystyle\left(\frac{_0\langle 1^1P_1|V_1|1^3D_1\rangle_0}
{E^0_{1^3D_1}-E^0_{1^1P_1}}\right)^2}}\;.
\label{C}
\end{eqnarray}
\end{widetext}

The detailed calculation of the matrix element $_0\langle1^1P_1|V_1|n^{(2s+1)}L_1\rangle_0$ is given in Appendix A.
Expanding these mixing coefficients up to $O(\delta^{}_c/m^{}_c)$ and with the results given in Eqs.\,(A15a) and (A15b), we obtain
\begin{widetext}
\bea                                          
&&\hspace{-0.5cm}C^{10}_{10}=1+O(\delta^2_c/m^2_c),~~~~~C^{11}_{10}(m^{}_f,m^{}_s)\equiv C^{11}_{10}\delta^{}_{m^{}_f,m^{}_s}=\frac{8}{3\sqrt{3}}\frac{\delta^{}_c}{m^{}_c}
\frac{I^{11}_{10}}{M^{}_{1P}-M^{}_{J/\psi}}\delta^{}_{m^{}_f,m^{}_s}+O(\delta^2_c/m^2_c),\nonumber\\
&&\hspace{-0.5cm}C^{20}_{20}=1+O(\delta^2_c/m^2_c),~~~~~C^{11}_{20}(m^{}_f,m^{}_s)\equiv C^{11}_{20}\delta^{}_{m^{}_f,m^{}_s}=-\frac{8}{3\sqrt{3}}\frac{\delta^{}_c}{m^{}_c}
\frac{I^{11}_{20}}{M^{}_{\psi'}-M^{}_{1P}}\delta^{}_{m^{}_f,m^{}_s}+O(\delta^2_c/m^2_c),\nonumber\\
&&\hspace{-0.5cm}C^{11}_{11}=1+O(\delta^2_c/m^2_c),~~~~~C^{10}_{11}(m^{}_f,m^{}_s)\equiv C^{10}_{11}\delta^{}_{m^{}_f,m^{}_s}=-\frac{8}{3\sqrt{3}}\frac{\delta^{}_c}{m^{}_c}
\frac{I^{10}_{11}}{M^{}_{1P}-M^{}_{J/\psi}}\delta^{}_{m^{}_f,m^{}_s}+O(\delta^2_c/m^2_c),\nonumber\\
&&\hspace{-0.5cm}C^{20}_{11}=1+O(\delta^2_c/m^2_c),~~~~~C^{12}_{11}(m^{}_f,m^{}_i+m^{}_s)\equiv C^{11}_{10}\delta^{}_{m^{}_f,m^{}_i+m^{}_s}=-\frac{8}{3}\sqrt{\frac{2}{3}}\frac{\delta^{}_c}{m^{}_c}
\frac{I^{12}_{11}}{M^{}_{1D}-M^{}_{1P}}\delta^{}_{m^{}_f,m^{}_i+m^{}_s}+O(\delta^2_c/m^2_c),\nonumber\\
&&\hspace{-0.5cm}C^{12}_{12}=1+O(\delta^2_c/m^2_c),~~~~~C^{11}_{12}(m^{}_f,m^{}_i+m^{}_s)\equiv C^{11}_{12}\delta^{}_{m^{}_f,m^{}_i+m^{}_s}=\frac{8}{3}\sqrt{\frac{2}{3}}\frac{\delta^{}_c}{m^{}_c}
\frac{I^{11}_{12}}{M^{}_{1D}-M^{}_{1P}}\delta^{}_{m^{}_f,m^{}_i+m^{}_s}+O(\delta^2_c/m^2_c),
\label{C'}
\eea
\end{widetext}
where $m^{}_i,m^{}_f, m^{}_s$ stands for the magnetic quantum numbers of the initial state orbital angular momentum, the final state orbital angular momentum, and the initial state spin, respectively. The numerical values of these mixing coefficients can be obtained once a potential model is chosen. In the CK model \cite{CK} and Cornell model \cite{Cornell}, the values are given in TABLE\,\ref{Cvalues} in APPENDIX B.

Actually, even for $\delta_c=0$, the states $|2^3S_1\rangle$ and $|1^3D_1\rangle$ are not just the
experimentally observed $|\psi'\rangle$ and $|\psi^{\prime\prime}\rangle$ since the leptonic width of
 $|1^3D_1\rangle$ is smaller than the experimentally measured value by an order of magnitude. Usually
 people believe that the observed $|\psi'\rangle$ and $|\psi^{\prime\prime}\rangle$ are mixtures of
 $|2^3S_1\rangle$ and $|1^3D_1\rangle$ \cite{KY90}\cite{Kuang02}:
 \begin{eqnarray}                      
 &&|\psi'\rangle=|2^3S_1\rangle\cos\theta~- |1^3D_1\rangle\sin\theta\nonumber\\
 &&|\psi^{\prime\prime}\rangle=|2^3S_1\rangle\sin\theta + |1^3D_1\rangle\cos\theta.
 \label{SDmixing}
 \end{eqnarray}
 The mixing angle $\theta$ can be determined by fitting the measured leptonic widths. The obtained
 values of $\theta$ in different models are \cite{KY90}\cite{Kuang02}
 \begin{eqnarray}                         
 &&{\rm Cornell~model}:~~~~~~~~~~~~~~~~\theta=10^\circ\nonumber\\
 &&{\rm Chen-Kuang~model}:~~~~~~\theta=12^\circ.
 \label{theta}
 \end{eqnarray}
Thus the physical $|\psi'\rangle$ and $|\psi^{\prime\prime}\rangle$ are [cf. Eq. (\ref{nmixing})]
\begin{eqnarray}                           
&&|\psi'\rangle=\cos\theta\bigg(C_{20}^{20}~|2^3S_1\rangle_0+C_{20}^{11}~|1^1P_1\rangle_0\bigg)\nonumber\\
&&~~~~-\sin\theta\bigg(C_{12}^{12}~|1^3D_1\rangle_0+C_{12}^{11}~|1^1P_1\rangle_0\bigg),\nonumber
\end{eqnarray}
\begin{eqnarray}
&&|\psi^{\prime\prime}\rangle=\sin\theta\bigg(C_{20}^{20}~|2^3S_1\rangle_0+C_{20}^{11}~|1^1P_1\rangle_0\bigg)\nonumber\\
&&~~~~+
\cos\theta\bigg(C_{12}^{12}~|1^3D_1\rangle_0+C_{12}^{11}~|1^1P_1\rangle_0
\bigg).
\label{psi'psi"}
\end{eqnarray}
So the hadronic transition $\psi'\to J/\psi+\pi^++\pi^-$ under consideration is expressed as
\begin{eqnarray}                         
&&\hspace{-1cm}\bigg[\cos\theta\bigg(C_{20}^{20}~|2^3S_1\rangle_0+C_{20}^{11}~|1^1P_1\rangle_0\bigg)\nonumber\\
&&~-\sin\theta\bigg(C_{12}^{12}~|1^3D_1\rangle_0+C_{12}^{11}~|1^1P_1\rangle_0\bigg)\bigg]\bigg|_{E=E_{20}}
\to\nonumber\\
&&~~\bigg(C_{10}^{10}~|1^3S_1\rangle_0+C_{10}^{11}~|1^1P_1\rangle_0\bigg)
\bigg|_{E=E_{10}}~+\pi+\pi.
\label{total transition}
\end{eqnarray}

\section{Calculation of the Transition Rates and Determination of the $\bm{O(\delta_c^1)}$ SPA Coefficients}
\subsection{$\bm{O(\delta_c^0)}$ and $\bm{O(\delta_c^1)}$ Transitions}

Hadronic transition process is dipicted in FIG.~\ref{HTfig}, in which the transition amplitude contains two factors,
namely the multipole gluon emission (MGE) factor and the hadronization (H) factor. We shall treat the two factors separately.

\begin{figure}[h]
\includegraphics[width=5.8truecm,clip=true]{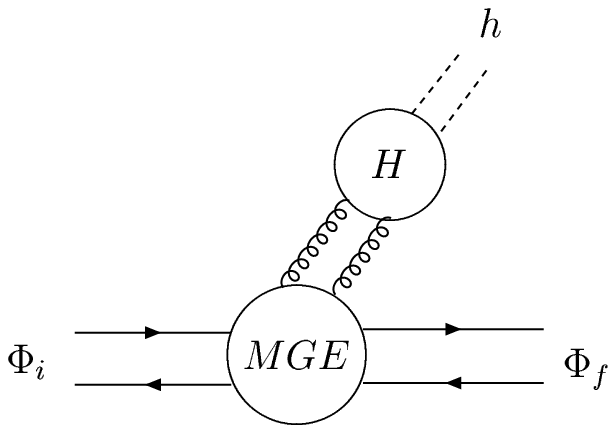}
\null\vspace{-0cm}
\caption{\footnotesize Diagram for a typical hadronic transition.}
\label{HTfig}
\end{figure}

We first consider the MGE part. Let $\bm E^a$ and $\bm B^a$ be the color-electric and color-magnetic fields, respectively.
In the conventional CP-conserving transitions, the MGE part contains certain quarkonium-gluon interaction vertices, e.g., the
color-electric dipole (E1) interaction, the color-magnetic dipole (M1) interaction, etc. \cite{Yan80,KY81,KYF,Kuang02}:
\bea                                           
&&\hspace{-0.8cm}{\rm E1}:-\bm d^{}_a\cdot \bm E^a(\bm X,t),\nonumber\\
&&\hspace{-0.6cm}\bm d^{}_a\equiv g^{}_E\int(\bm x-\bm X)\Psi^\dagger(\bm x,t)
\frac{\lambda_a}{2}\Psi(\bm x,t)d^3x,\nonumber\\
&&\hspace{-0.8cm}{\rm M1}:\bm m^{}_a\cdot \bm B^a(\bm X,t),\nonumber\\
&&\hspace{-0.6cm}\bm m^{}_a\equiv \frac{g^{}_M}{2}\int (\bm x-\bm X)\times\Psi^\dagger(\bm x,t)\frac{\lambda_a}{2}\gamma^0\bm{\gamma}\Psi(\bm x,t)d^3x,
\label{E1,M1}
\eea
where $g^{}_E$ and $g^{}_M$ are effective color-electric and color-magnetic coupling strengths, respectively, $\bm X$ is the center-of-mass coordinate of the quarkonium, $\Psi(\bm x,t)$ is the quarkonium wave function at the space-time
point $(\bm x,t)$.
For the CP-violating CEDM interaction vertex. Eq.~(\ref{cEDM}) can be written as
\begin{eqnarray}                              
{\cal L}^{}_{CEDM}&=&id'_c\psi^\dagger_c\gamma^0\sigma^{0k}\gamma_5\frac{\lambda_a}{2}\psi_cG^a_{0k}
-id'_c\psi^\dagger_c\gamma^0\sigma^{ij}\gamma_5\frac{\lambda_a}{2}\psi_cG_{ij}\nonumber\\
&=&-\frac{\delta_c}{m_c}\psi^\dagger_c
\begin{pmatrix}
\sigma_i~~~~0\\ 0~~-\sigma_i
\end{pmatrix}
\frac{\lambda_a}{2}\psi_cE^a_i\nonumber\\
&&-i\frac{\delta_c}{m_c}\psi^\dagger_c
\begin{pmatrix}
0~~~~\sigma_i\\-\sigma_i~~0
\end{pmatrix}
\frac{\lambda_a}{2}\psi_cB^a_i.
\label{c-EDMinteraction}
\end{eqnarray}
The first term is just the interaction between the CEDM and the color-electric field.
The second term is off diagonal, so that it is nonvanishing only when the lower components of the
quark spinor $\displaystyle{\pm\frac{\bm{p\cdot\sigma}}{E+m}}$ are taken into account, i.e., when the
quark is moving. So the second term means the interaction between the color current and the color
magnetic field. In the nonrelativistic limit, only the first term is novanishing, and the interaction
vertex is:
\begin{eqnarray}                          
{\rm CEDM}:~~~~-i\frac{\delta_c}{m_c}(\frac{\lambda_a}{2}\bm\sigma+\frac{\bar\lambda_a}{2}\bar{\bm\sigma})
\cdot\bm{E}^a.
\label{EDMvertex}
\end{eqnarray}
As in the case of M1 transition given in Ref.~\cite{Yan80}, after certain treatment of the color
factor, the effective CEDM vertex is
\begin{eqnarray}                          
{\rm CEDM}:~~~~-i\frac{\delta_c}{2m_c}(\bm{\sigma}-\bar{\bm \sigma})
\cdot\bm{E},
\label{EDMvertex'}
\end{eqnarray}
where $\bm E\equiv \frac{\lambda^{}_a}{2}\bm E^a_{}$.

With these vertices, the transition (\ref{total transition}) can be divided into the $O(\delta_c^0)$ terms and $O(\delta_c^1)$ terms. They are:\\
\begin{widetext}
\null\noindent
{\bf (i) $\bm{O(\delta_c^0)}$ term}:\\

(a) Ordinary E1-E1 transition of $2S\to 1S$ with the coefficients
$\cos\theta C_{20}^{20}C_{10}^{10}$:
\begin{eqnarray}                          
\cos\theta C_{20}^{20}~|2^3S_1\rangle_0\bigg|_{E=E_{20}}\to C_{10}^{10}|1^3S_1\rangle_0\bigg|_{E=E_{10}}
+\pi+\pi.
\label{2S-1Spipi}
\end{eqnarray}

(b) Ordinary E1-E1 transition of $1D\to 1S$ with the coefficient $-\sin\theta C^{12}_{12}C^{10}_{10}$:
\begin{eqnarray}                           
-\sin\theta C_{12}^{12}~|1^3D_1\rangle_0\bigg|_{E=E_{20}}\to C_{10}^{10}~|1^3S_1\rangle_0
\bigg|_{E=E_{10}}~+\pi+\pi.
\label{1D-1Spipi}
\end{eqnarray}
{\bf (ii) $\bm{O(\delta_c^1)}$ terms}:\\

(c) Ordinary E1-M1 transition of $2S\to 1P$ with the coefficients
$\cos\theta C_{20}^{20}C_{10}^{11}$:
\begin{eqnarray}                       
\cos\theta C_{20}^{20}~|2^3S_1\rangle_0\bigg|_{E=E_{20}}\to C_{10}^{11}~|1^1P_1\rangle_0\bigg|_{E=
E_{10}}~+\pi+\pi.
\label{1S-1Ppipi}
\end{eqnarray}

(d) Ordinary E1-M1 transition of $1D\to 1P$ with the coefficients
$-\sin\theta C_{12}^{12}C_{10}^{11}$:
\begin{eqnarray}                      
-\sin\theta C_{12}^{12}~|1^3D_1\rangle_0\bigg|_{E=E_{20}}\to C_{10}^{11}~|1^1P_1\rangle_0
\bigg|_{E=E_{10}}+\pi+\pi.
\label{1D-1Ppipi}
\end{eqnarray}

(e) Ordinary $E1-M1$ transition of $1P\to 1S$ with the coefficients
$(\cos\theta C_{20}^{11}-\sin\theta C_{12}^{11})$:
\begin{eqnarray}                       
(\cos\theta C_{20}^{11}-\sin\theta C_{12}^{11})~|1^1P_1\rangle_0\bigg|_{E=E_{20}}\to
C_{10}^{10}~|1^3S_1\rangle_0\bigg|_{E=E_{10}}+\pi+\pi.
\label{1P-1S}
\end{eqnarray}

(f) M1-CEDM1 transition of $2S\to 1S$ with the coefficients
$\cos\theta C_{20}^{20}C_{10}^{10}$
\begin{eqnarray}                         
\cos\theta C_{20}^{20}~|2^3S_1\rangle_0\bigg|_{E=E_{20}}\to
C_{10}^{10}~|1^3S_1\rangle_0\bigg|_{E=E_{10}}+\pi+\pi.
\label{2S-1SEDM}
\end{eqnarray}

(g) E1-CEDM2 transitions of $2S\to 1S$ and $1D\to 1S$ with coefficients $\cos\theta C^{20}_{20}C^{10}_{10}$ and $-\sin\theta C_{12}^{12}C_{10}^{10}$, respectively:
\begin{eqnarray}                            
\cos\theta C_{20}^{20}~|2^3S_1\rangle_0\bigg|_{E=E_{20}}\to
C_{10}^{10}~|1^3S_1\rangle_0\bigg|_{E=E_{10}}+\pi+\pi.
\label{2S-1S E1-CEDM2}
\end{eqnarray}
\begin{eqnarray}                            
-\sin\theta C_{12}^{12}~|1^3D_1\rangle_0\bigg|_{E=E_{20}}\to
C_{10}^{10}~|1^3S_1\rangle_0\bigg|_{E=E_{10}}+\pi+\pi.
\label{1D-1S E1-CEDM2}
\end{eqnarray}
\end{widetext}

For a given potential model, there is a systematic way of calculating the MGE factors \cite{KY81}\cite{Kuang06}.

Next we consider the hadronization (H) part, the matrix elements reflecting the conversion of gluons into light hadrons.
These matrix elements are at the scale of a few hundred MeV, and the calculation is thus highly nonperturbative.
So far, there is no reliable way of calculating them from the first principles of QCD, so that we have to take certain
phenomenological approach. A conventionally used approach is the soft pion approach (SPA) in which the H-factor matrix
element is phenomenologically expressed in terms of power expansion of the momenta of the two pions with unknown coefficients
\cite{BrownCahn}. To lowest nonvanishing order, the number of unknown coefficients is usually not large, so that they can be
determined from taking certain experimental input data. This approach has proved to be successful in calculating the transition rates,
the $M_{\pi\pi}$ distributions, etc. \cite{KY81}\cite{Kuang06}. However, in the present case, it will be difficult if we merely
take this approach since there are so many kinds of H-factor matrix elements listed in (\ref{2S-1Spipi})$\--$(\ref{1D-1S E1-CEDM2})
containing too many unknown coefficients. There are not enough known input experimental data to determine them. Another viable but
cruder approach is the two-gluon approach (2GA) proposed and used in Refs.~\cite{KY81} and \cite{Kuang06}. In this approach, the
external pion-fields in the H-factor are approximately replaced by two external gluons, so that the matrix elements can be easily
evaluated as functions of the pion- (gluon-) momenta and the two phenomenological coupling constants $g^{}_E$ and $g^{}_M$ which can be
determined by known experimental inputs. Of course this is a crude order of magnitude estimate. However, it has been shown that
this crude approach does give right order of magnitudes of transition rates for many processes \cite{KY81}\cite{Kuang06}\cite{KTY88}. 
The shortcoming of the 2GA is that it cannot give the correct angular-dependent distributions such as angular distributions, $M_{\pi\pi}$
distributions, etc. This is because that the pion is spinless, while the gluon spin is 1. In this situation, we shall take both SPA and
2GA in this paper for complementarity.

We can first take the 2GA to calculate all the transition rates listed in (\ref{2S-1Spipi})$\--$(\ref{1D-1S E1-CEDM2}) which contain
two effective coupling constants $g^{}_E$ and $g^{}_M$. Since the total transition rate $\Gamma(\psi'\to J/\psi~\pi\pi)$ is essentially
contributed by the $O(\delta_c^0)$ rate which only contains $g^{}_E$, we can take the experimental value of $\Gamma(\psi'\to J/\psi~\pi\pi)$ \cite{PDG},
\begin{eqnarray}                                        
&&\Gamma^{}_{\psi'}=304\pm9~{\rm keV},\nonumber\\
&&B(\psi'\to J/\psi\,\pi^+\pi^-)=(33.6\pm0.4)\%,\nonumber\\
&&B(\psi'\to J/\psi\,\pi^0\pi^0)=(17.73\pm0.34)\%,
\label{GammaExpt}
\end{eqnarray}
as input to determine $g^{}_E$. For the determination of $g^{}_M$, we can take the 2GA calculated branching ratio
$B(\psi'\to h_c\pi^0)\times B(h_c\to\eta_c\gamma)$ \cite{Kuang02} containing $g^{}_M/g^{}_E$ to compare with the corresponding experimental value
\cite{BESIII-h_c}. For example, in the CK model and the Cornell model, the determined $g^{}_E$ and $g^{}_M/g^{}_E$ are:
\bea
&&\null\noindent{\rm\bf CK~model}: \nonumber\\                          
&&\null\noindent~~~~\alpha^{}_E\equiv\frac{g_E^2}{4\pi}=0.523,~~
\frac{\alpha^{}_M}{\alpha^{}_E}=\frac{g_M^2}{g_E^2}=2.36,\nonumber\\
&&\null\noindent{\rm\bf Cornell~model}:\nonumber\\
&&\null\noindent~~~~\alpha^{}_E\equiv\frac{g_E^2}{4\pi}=0.667,~~\frac{\alpha^{}_M}{\alpha^{}_E}=\frac{g_M^2}{g_E^2}=2.36.
\label{alpha_E,alpha_M}
\eea
With the determined $g^{}_E$ and $g^{}_M$, we can obtain all the relative sizes of the transition rates
in (\ref{2S-1Spipi})$\--$(\ref{1D-1S E1-CEDM2}) with only one undetermined parameter $d'_c$ (or $\delta_c$) left in the $O(\delta_c^1)$ rates.
Next, we can take the SPA to calculate all the 
transition rates containing certain unknown coefficients. As a perturbation calculation, we first obtain the $O(\delta_c^0)$ transition rate containing three unknown coefficients \cite{BrownCahn} which can be determined by the data (\ref{GammaExpt}) and $M^{}_{\pi\pi}$ distribution as what is conventionally done \cite{CLEO84}. Then we calculate the $O(\delta_c^1)$ transition rates containing several new unknown coefficients.
Comparing the SPA and 2GA results, we can express these unknown coefficients in terms of the known $O(\delta_c^0)$ coefficients, so the SPA results of the $O(\delta_c^1)$ transition amplitudes contain only one undetermined parameter $d'_c$ (or $\delta_c$). We can then calculate the total distribution
$d\Gamma(\psi'\to J/\psi+\pi^++\pi^-)/dM_{\pi\pi}$ to compare with the BES data for testing $d'_c$ (or $\delta_c$). 
We shall see that the inclusion of the $O(\delta_C^1)$ contributions does improve the fit, and the best fit value of $\delta_c$ is nonvanishing. However, considering the experimental errors, $\delta_c$ is still consistent with zero. So we can obtain an upper bound on $\delta_c$ (or $d'_c$). The detailed analysis of this kind of study will be given in Sec. IV.\\

\subsection{Transition Rates of the $\bm{O(\delta_c^0)}$ processes (\ref{2S-1Spipi}) and (\ref{1D-1Spipi})}

The $O(\delta_c^0)$ transition rates in Eqs.~(\ref{2S-1Spipi}) and (\ref{1D-1Spipi}) have been calculated in the published papers \cite{KY81}\cite{Kuang06}\cite{KY90}. Here we list the results.

The transitions in (\ref{2S-1Spipi})---(\ref{1D-1Spipi}) belong to the ordinary E1E1 transitions.
The E1E1 transition amplitude can be written as \cite{KY81}\cite{Kuang06}
\begin{widetext}
\begin{eqnarray}                            
{\cal M}_{E1E1}&=&i\frac{g_E^2}{6}\sum_{KL}\frac{\langle\Phi^{}_F|r^{}_{m^{}_1}|KL\rangle
\langle KL|r^{}_{m^{}_2}|\Phi^{}_I\rangle}
{E_I-E_{KL}}\langle \pi\pi|E^a_{-m^{}_1} E^a_{-m^{}_2}|0\rangle\nonumber\\
&=&i\frac{g_E^2}{6}f^{111}_{n^{}_Il^{}_In^{}_Fl^{}_F}\langle l^{}_Fs^{}_Fj^{}_Fm^{}_F|\hat r^{}_{m^{}_1}\hat r^{}_{m^{}_2}|l^{}_Is^{}_Ij^{}_Im^{}_I\rangle
\langle \pi\pi|E^a_{-m^{}_1} E^a_{m^{}_2}|0\rangle,
\label{E1E1amplitude}
\end{eqnarray}
\end{widetext}
where $\Phi^{}_I~(\Phi^{}_F)$ is the initial (final) quarkonium state, $\bm r$ is the separation vector between the two heavy quarks, $\hat r$ is the unit vector of $\bm r$, $r^{}_{m_1}\,(r^{}_{m_2})$ is the component of $\bm r$ in the spherical coordinate system with the magnetic quantum number $m^{}_{m_1}\,(m^{}_{m_2})\in \{1,0,-1\}$ (cf. Eq.\,(A3) in APPENDIX A), $l,s.j.m$ are, respectively, the orbital angular momentum, the spin, the total angular momentum, and the total magnetic quantum numbers of the quarkonium state, $K$ ($L$) is the principal (orbital angular momentum) quantum number of the intermediate state, and $E_{KL}$ is the energy eigenvalue of the intermediate vibrational state $|KL\rangle$. The factor $\langle l^{}_Fs^{}_Fj^{}_Fm^{}_F|\hat r^{}_k\hat r^{}_l|l^{}_Is^{}_Ij^{}_Im^{}_I\rangle$ can be evaluated using the properties of the spherical harmonics \cite{KY81}, and the reduced amplitude $f^{111}_{n^{}_Il^{}_In^{}_Fl^{}_F}$ is
\begin{widetext}
\begin{eqnarray}                     
f^{LP^{}_IP^{}_F}_{n^{}_Il^{}_In^{}_Fl^{}_F}\equiv \sum_{K}\frac{\int R_F(r)r^{P_F}R^*_{KL}(r)r^2dr
\int R^*_{KL}(r^\prime)
r^{\prime P_I}R_I(r^\prime)r^{\prime 2}dr^\prime}{M_I-E_{KL}},
\label{f}
\end{eqnarray}
\end{widetext}
in which $R_I$, $R_F$, and $R_{KL}$ are radial wave functions of
the initial, final, and intermediate vibrational states,
respectively. These radial wave functions are calculated from the
Schr\"odinger equation with a given potential model. The values of various $f^{LP^{}_IP^{}_F}_{n^{}_Il^{}_In^{}_Fl^{}_F}$ in the CK and Cornell models are listed in TABLE\,\ref{fvalues} in APPENDIX\,B.

In the following, we consider the approaches to the hadronization factor $\langle \pi\pi|E^a_k E^a_l|0\rangle$.

\null\noindent
{\bf i. SPA}

In the SPA, the hadronization  factor can be generally parametrized as \cite{BrownCahn}
\begin{widetext}
\begin{eqnarray}                    
\null\hspace{-0.4cm}&&\frac{g_E^2}{6}\langle\pi_\alpha(q_1)\pi_\beta(q_2)|E^a_{-m^{}_1} E^a_{-m^{}_2}|0\rangle
=\frac{\delta_{\alpha\beta}}{\sqrt{(2\omega_1)(2\omega_2)}}
\bigg\{\delta_{m^{}_1m^{}_2}[{\cal A}q^\mu_1 q^{}_{2\mu}
+{\cal B}\omega^{}_1\omega^{}_2]
+{\cal C}\bigg(q^{}_{1m^{}_1}q^{}_{2m^{}_2}+q^{}_{1m^{}_2}q^{}_{2m^{}_1}-\frac{2}{3}\delta_{m^{}_1m^{}_2}{\bm q}_1
\cdot{\bm q}_2\bigg)\bigg\}\nonumber\\
&&\null\hspace{1.8cm}=\frac{{\cal A}\delta^{}_{\alpha\beta}}{\sqrt{2\omega^{}_1 2\omega^{}_2}}\bigg\{\delta^{}_{m^{}_1m^{}_2}[q^\mu_1 q^{}_{2\mu}
+\frac{\cal B}{\cal A}\omega^{}_1\omega^{}_2]
+\frac{\cal C}{\cal A}\bigg(q^{}_{1m^{}_1}q^{}_{2m^{}_2}+q^{}_{1m^{}_2}q^{}_{2m^{}_1}-\frac{2}{3}\delta_{m^{}_1m^{}_2}{\bm q}_1
\cdot{\bm q}_2\bigg)\bigg\},
\label{HofE1E1}
\end{eqnarray}
\end{widetext}
where $\cal A,B$ and $\cal C$ are phenomenological constants, and $q^{}_\alpha=(\omega^{}_\alpha,\bm q^{}_\alpha)$ is the four momentum of $\pi^{}_\alpha$.
For a given invariant mass $M^{}_{\pi\pi}$, the {\cal A} term 
is angular-independent, while the ${\cal B}$ and ${\cal C}$ terms 
are angular-dependent \cite{BrownCahn}.

For $|2^3S_1\rangle\to |1^3S_1\rangle\pi\pi$, the the main contributions to the total transition rate are from the ${\cal A}$ and ${\cal B}$ terms \cite{CLEO84}, while for $|1^3D_1\rangle\to |1^3S_1\rangle\pi\pi$, the main contribution is from the ${\cal C}$ term \cite{KY81}. Thus
the $M^{}_{\pi\pi}$ distribution is \cite{KY81} 
\begin{widetext}
\bea                                  
&&\hspace{-0.8cm}\frac{d\Gamma(\psi'\to J/\psi\,\pi\pi)^{}_{SPA}}{dM^{}_{\pi\pi}}=|{\cal A}|^2\bigg[\cos^2\theta ~\frac{dG^{}_{\cal AB}(\psi^\prime)}{dM^{}_{\pi\pi}}
~|f^{111}_{2010}(\psi^\prime)|^2
+\sin^2\theta ~\frac{dH^{}_{\cal C}(\psi^{\prime})}{dM^{}_{\pi\pi}}
~|f^{111}_{1210}(\psi^{\prime})|^2\bigg]\nonumber\\
&&\hspace{-0.4cm}=|{\cal A}|^2\cos^2\theta\frac{M^{}_{\pi\pi}}{4\pi^3}\frac{M^{}_{J/\psi}}{M^{}_{\psi'}}
\bigg\{\frac{1}{4}(M_{\pi\pi}^2-2m_\pi^2)^2{\cal F}^{}_0
+\frac{\cal B}{\cal A}(M_{\pi\pi}^2-2m_\pi^2){\cal F}^{}_1+\bigg|\frac{\cal B}{\cal A}\bigg|^2{\cal F}^{}_2\bigg\}|f^{111}_{2010}(\psi^\prime)|^2\nonumber\\
&&\hspace{-0.4cm}~~+\sin^2\theta\frac{M^{}_{\pi\pi}}{225\pi^3}\frac{M^{}_{J/\psi}}{M^{}_{\psi'}}\bigg|\frac{\cal C}{\cal A}\bigg|^2\bigg\{\bigg[3m_\pi^2(m_\pi^2-K_0^2)+\frac{1}{4}(M_{\pi\pi}^2-2m_\pi^2)^2\bigg]{\cal F}^{}_0+(8m_\pi^2-M_{\pi\pi}^2){\cal F}^{}_1+4{\cal F}^{}_2
\bigg\}|f^{111}_{1210}(\psi^{\prime})|^2,
\label{dGamma/dM_pipi}
\eea
where
\begin{eqnarray}                      
&&\null\hspace{-0.8cm}{\cal F}^{}_0=\frac{|\bm K|}{M_{\pi\pi}}\sqrt{M_{\pi\pi}^2-4m_\pi^2},\hspace{3cm}
{\cal F}^{}_1={\cal F}^{}_0\bigg[\frac{1}{6}\frac{K_0^2}{M_{\pi\pi}^2}(M_{\pi\pi}^2+2m_\pi^2)+\frac{1}{12}(M_{\pi\pi}^2-4m_\pi^2)\bigg],
\nonumber\\
&&\null\hspace{-0.8cm}{\cal F}^{}_2=\frac{{\cal F}^{}_0}{80}\bigg[\frac{8}{3}\frac{K_0^4}{M_{\pi\pi}^4}(M_{\pi\pi}^4+2M_{\pi\pi}^2m_\pi^2+6m_\pi^4)
+\frac{4}{3}\frac{K_0^2}{M_{\pi\pi}^2}(M_{\pi\pi}^2-4m_\pi^2)(M_{\pi\pi}^2+6m_\pi^2)+(M_{\pi\pi}^2-4m_\pi^2)^2\bigg],
\label{I}
\end{eqnarray}
in which $(K_0,\bm K)$ is the four momentum of the $\pi\pi$ system,
\begin{eqnarray}                                    
&&\hspace{-0.9cm}|{\bm K}|\equiv\frac{1}{2M^{}_{\psi'}}\left\{[(M^{}_{\psi'}+M^{}_{\pi\pi})^2-M_{J/\psi}^2][(M^{}_{\psi'}-M^{}_{\pi\pi})^2-M_{J/\psi}^2]\right\}^{1/2},~~~~~~
K_0\equiv\frac{1}{M^{}_{\psi'}}\left(M_{\psi'}^2+M_{\pi\pi}^2-M_{J/\psi}^2\right).
\label{K,K_0}
\end{eqnarray}

 The determination of $\cal A,~B/A$ and $\cal C/A$ will be discussed in Sec.~IV.

Integrating over $dM^{}_{\pi\pi}$ in (\ref{dGamma/dM_pipi}), we obtain the transition rate
\begin{eqnarray}                       
\Gamma\big(\psi^\prime\to J/\psi~\pi\pi\big)^{}_{SPA}=|{\cal A}|^2\bigg[\cos^2\theta ~G^{}_{\cal AB}(\psi^\prime)
~|f^{111}_{2010}(\psi^\prime)|^2
+\sin^2\theta ~H^{}_{\cal C}(\psi^{\prime})
~|f^{111}_{1210}(\psi^{\prime})|^2\bigg],
\label{psi'rate}
\end{eqnarray}
where
\begin{eqnarray}                      
&&\null\hspace{-0.5cm}G^{}_{\cal AB}\equiv\frac{1}{8\pi^3}\frac{M^{}_{J/\psi}}{M^{}_{\psi'}}\int^{\Delta M}_{2m^{}_\pi} dM^{2}_{\pi\pi}\bigg\{\frac{1}{4}(M_{\pi\pi}^2-2m_\pi^2)^2{\cal F}^{}_0
+\frac{\cal B}{\cal A}(M_{\pi\pi}^2-2m_\pi^2){\cal F}^{}_1+\bigg|\frac{\cal B}{\cal A}\bigg|^2{\cal F}^{}_2\bigg\},
\nonumber\\
&&\null\hspace{-0.5cm}H^{}_{\cal C}=\frac{1}{450\pi^3}\frac{M^{}_{J/\psi}}{M^{}_{\psi'}}\int^{\Delta M}_{2m^{}_\pi} dM^{2}_{\pi\pi}\bigg|\frac{\cal C}{\cal A}\bigg|^2\bigg\{\bigg[3m_\pi^2(m_\pi^2-K_0^2)+\frac{1}{4}(M_{\pi\pi}^2-2m_\pi^2)^2\bigg]{\cal F}^{}_0+(8m_\pi^2-M_{\pi\pi}^2){\cal F}^{}_1+4{\cal F}^{}_2
\bigg\},
\label{G,H}
\end{eqnarray}
\end{widetext}
in which $\Delta M\equiv M^{}_{\psi'}-M^{}_{J/\psi}$,

Although this approach can give the reliable $M_{\pi\pi}^{}$ distribution, it cannot be applied to other processes in (\ref{2S-1Spipi})$\--$(\ref{1D-1S E1-CEDM2}) since other processes contain new unknown parameters in their hadronization factors, and there are not enough experimental data to determine them. So we have to take the help of the 2GA.

\null\noindent
{\bf ii. 2GA}

The hadronization factor can be further written as
\bea                                    
\langle\pi\pi|E^a_{-m^{}_1}E^a_{-m^{}_2}|0\rangle=\sum_N \langle\pi\pi|N\rangle\langle N|E^a_{-m^{}_1}E^a_{-m^{}_2}|0\rangle,
\label{insert |N>}
\eea
where $|N\rangle$ denotes a complete set of intermediate states. The 2GA assumes that the color-singlet 2-gluon state $|N\rangle=|g^cg^c\rangle$ dominates. In general, the factor $\langle\pi\pi|g^cg^c\rangle$ is a function of the pion momenta. Considering the fact that the
hadronnization factors in most hadronic transition processes are all at the few hundred MeV scale, the running of $\langle\pi\pi|g^cg^c\rangle$
with the momentum in such a range of scale is mild. So that we approximately take $\langle\pi\pi|g^cg^c\rangle\approx const.$, and the constant can be absorbed into the redefinition of the effective coupling constant $g^{}_E$. Thus we have
\bea                                    
 \langle\pi\pi|E^a_{-m^{}_1}E^a_{-m^{}_2}|0\rangle\approx\langle g^cg^c|E^a_{-m^{}_1}E^a_{-m^{}_2}|0\rangle.
\label{2GA}
\eea
This approximation can be extended to other hadronization factors containing color magnetic field(s).

The matrix element $\langle g^cg^c|E^a_{-m^{}_1}E^a_{-m^{}_2}|0\rangle$ can be easily evaluated:
\begin{widetext}
\begin{eqnarray}                                  
\null\hspace{-0.3cm}\langle g^cg^c|E^a_{-m^{}_1}E^b_{-m^{}_2}|0\rangle
=-\frac{\delta^{}_{ab}}{\sqrt{2\omega^{}_1}\sqrt{2\omega^{}_2}}\bigg[\omega^{}_1\epsilon_{1,-m^{}_1}(\lambda^{}_1)
\omega^{}_2\epsilon^{}_{2,-m^{}_2}(\lambda^{}_2)
+\omega^{}_1\epsilon^{}_{1,-m^{}_2}(\lambda^{}_1)\omega^{}_2\epsilon^{}_{2,-m^{}_1}(\lambda^{}_2)\bigg],
\label{2-gluon}
\end{eqnarray}
\end{widetext}
where $\omega~(\omega')$ and $\bm\epsilon~(\bm\epsilon')$ are the energy and polarization vector of the gluon, respectively. After lengthy but elementary calculations, we obtain
\begin{widetext}

\bea                                   
&&\Gamma^{}_{E1E1}(\psi'\to J/\psi~\pi\pi)^{}_{2GA}=\left(\frac{g_E^2}{6}\right)^2\frac{8}{27\pi^3}\frac{(\Delta M)^7}{140}\bigg[\cos^2\theta|f^{111}_{2010}|^2
+\frac{2}{5}\sin^2\theta|f^{111}_{1210}|^2\bigg].
\label{2GA_Gamma(psi'-psi)}
\eea
\end{widetext}

In the following we shall take the 2GA to calculate other transition rates in (\ref{2S-1Spipi})$\--$(\ref{1D-1S E1-CEDM2}), and compare them with (\ref{2GA_Gamma(psi'-psi)}) to determine their relative sizes.

\subsection{Transition Rates of the $\bm{O(\delta_c^1)}$ E1M1 processes (\ref{1S-1Ppipi}), (\ref{1D-1Ppipi}) and (\ref{1P-1S})}

These processes contain $\delta_c$ through the mixing coefficients $C^{11}_{20}$, $C^{11}_{10}$,
and $(\cos\theta C^{11}_{20}-\sin\theta C^{11}_{12})$, respectively [cf. Eqs\,(\ref{C})]. The transition amplitudes are $|2^3S_1\rangle\to|1^1P_1\rangle+\pi^++\pi^-,\,
|1^1P_1\rangle\to|1^3S_1\rangle+\pi^++\pi^- $ and $|1^3D_1\rangle\to|1^1P_1\rangle+\pi^++\pi^-$. They all belong to the ordinary E1M1 transitions which have been calculated in Refs.\,\cite{KY81}\cite{Kuang02}\cite{KTY88}.  Here we present the results as follows.

For these three processes, let us denote the initial and final quarkonium states by $\Phi^{}_I$ and $\Phi^{}_F$, and the spins of $c$ and $\bar{c}$ by $\bm s_c$ and $\bm s_{\bar c}$.
Then the  E1M1 transition amplitude can be generally written as
\begin{widetext}
\begin{eqnarray}               
{\cal M}_{E1M1}&=&i\frac{g^{}_Eg^{}_M}{12m^{}_c}\sum_{KLM^{}_s}\frac{\langle \Phi^{}_F|r_{m^{}_1}|KLM^{}_s\rangle
\langle KLM^{}_s|(s_c-s_{\bar c})_{m^{}_2}|\Phi^{}_I\rangle+
\langle \Phi^{}_F|(s_c-s_{\bar c})_{m^{}_2}|KLM^{}_s\rangle
\langle KLM^{}_s|r_{m^{}_1}|\Phi^{}_I\rangle}{M_{\psi^\prime}-E_{KL}}\nonumber\\
&&\times\langle \pi^+\pi^-|E_{-m^{}_1}B_{-m^{}_2}|0\rangle.
\label{E1M1amplitude}
\end{eqnarray}
\end{widetext}
Taking the technique of evaluating the spin matrix elements in the MGE factors (cf. Appendix), and after certain lengthy calculations, we obtain

\null\vspace{0.2cm}\noindent
{\bf (a)} $\bm{|2^3S_1\rangle\to|1^1P_1\rangle+\pi+\pi}$
\bea                                   
&&\null\hspace{-0.5cm}{\cal M}_{E1M1}(2^3S_1\to 1^1P_1\pi\pi)^{}=\frac{\delta^{}_{ab}g^{}_Eg^{}_M}{6\sqrt{3}m^{}_c}(f^{110}_{2011}+f^{001}_{2011})\nonumber\\
&&\hspace{0cm}\times\sum_{m_1m_2}\{a^\ast_{m_1}(m_F^{})a_{m_2}(m^{}_I)\}\langle\pi\pi|E^{}_{-m_1}B^{}_{-m_2}|0\rangle,
\label{E1M1-2S-1P}
\eea
where the definitions of $a^\ast$ and $a$ are given in Eq.\,(A5), $m^{}_I$ and $m^{}_F$ are the magnetic quantum numbers of the initial- and final-state, and the definition of $f^{LP^{}_IP^{}_F}_{n^{}_Il^{}_In^{}_Fl^{}_F}$ is given in Eq.\,(\ref{f}).

\null\vspace{0.2cm}\noindent
{\bf (b)} $\bm{|1^1P_1\rangle\to|1^3S_1\rangle+\pi+\pi}$
\bea                                   
&&\null\hspace{-0.5cm}{\cal M}_{E1M1}(1^1P_1\to 1^3S_1\pi\pi)^{}=\frac{\delta^{}_{ab}g^{}_Eg^{}_M}{6\sqrt{3}m^{}_c}(f^{010}_{1110}+f^{101}_{1110})\nonumber\\
&&\hspace{0cm}\times\sum_{m_1m_2}\{a^\ast_{m_1}(m_F^{})a_{m_2}(m^{}_I)\}\langle\pi\pi|E^{}_{-m_1}B^{}_{-m_2}|0\rangle.
\label{E1M1-1P-1S}
\eea
and
\begin{widetext}
\null\vspace{0.2cm}\noindent
{\bf (c)} $\bm{|1^3D_1\rangle\to|1^1P_1\rangle+\pi+\pi}$
\bea                                   
&&\null\hspace{-0.5cm}{\cal M}_{E1M1}(1^3D_1\to 1^1P_1\pi\pi)^{}=\frac{\delta^{}_{ab}g^{}_Eg^{}_M}{\sqrt{6}m^{}_c}(f^{110}_{1211}+f^{201}_{1211})
\sum_{m_1m_2}\{a^\ast_{m_1}(m_F^{})a_{m_2}(m^{}_I)\}\langle\pi\pi|E^{}_{-m_1}B^{}_{-m_2}|0\rangle.
\label{E1M1-1D-1P}
\eea

Next we take the SPA and 2GA to evaluate the hadronization factor $\langle\pi\pi|E^{}_{-m_1}B^{}_{-m_2}|0\rangle$.

\null\noindent
{\bf i. SPA}

The hadronization factor $\langle\pi\pi|E^{}_{-m_1}B^{}_{-m_2}|0\rangle$ is a second rank pseudo-tensor which is to be expressed in terms of the pion momenta in the SPA. To lowest order, the expression is of the form
\bea                                    
\hspace{-0.5cm}g^{}_Eg^{}_M\langle\pi\pi|E^{}_{-m_1}B^{}_{-m_2}|0\rangle&=&{\cal K}^{}_{E1M1}\epsilon^{}_{-m_1,-m_2,-m_3}
\frac{\omega^{}_1 q^{}_{2,m^{}_3}+\omega^{}_2 q^{}_{1,m^{}_3}}{\sqrt{2\omega 2\omega'}},
\label{E1M1-SPA}
\eea
where ${\cal K}^{}_{E1M1}$ is a phenomenological constant. With this expression for the hadronization factor, the amplitudes (\ref{E1M1-2S-1P})$\--$(\ref{E1M1-1D-1P}), after certain evaluation, are
\begin{eqnarray}                         
\hspace{-0.9cm}\mathcal{M}^{}_{E1M1}(2^3S_1\to 1^1P_1)&=&\frac{\sqrt{2}{\cal K}^{}_{E1M1}\delta_{ab}}{\sqrt{2\omega_12\omega_2}}\cos\theta C^{11}_{10}(f^{110}_{2011}+f^{001}_{2011})\nonumber\\
&&\times(-1)^{1+m^{}_f}\left(
                                                                \begin{array}{ccc}
                                                                  1 & 1 & 1 \\
                                                                  m^{}_i+m^{}_s & -m & -m^{}_f \\
                                                                \end{array}
                                                              \right)
(q^{}_1\omega^{}_2+q^{}_2\omega^{}_1)^{}_{m},~~
\label{SPA-E1M1-2S-1P}
\end{eqnarray}
\begin{eqnarray}                      
\hspace{-0.8cm}\mathcal{M}^{}_{E1M1}(1^1P_1\to 1^3S_1)&=&-\frac{\sqrt{2}{\cal K}^{}_{E1M1}\delta_{ab}}{\sqrt{2\omega_12\omega_2}}\sin\theta C^{11}_{10}(f^{010}_{1110}+f^{101}_{1110})\nonumber\\
&&\times(-1)^{m^{}_f+m^{}_s}\left(
                                                                \begin{array}{ccc}
                                                                  1 & 1 & 1 \\
                                                                  m^{}_i & -m & -m^{}_f-m^{}_s \\
                                                                \end{array}
                                                              \right)
(q^{}_1\omega^{}_2+q^{}_2\omega^{}_1)^{}_{m},~~~~
\label{SPA-E1M1-1P-1S}
\end{eqnarray}
\begin{eqnarray}                      
\hspace{-0.9cm}\mathcal
{M}^{}_{E1M1}(1^3D_1\to 1^1P_1)&=&\frac{{\cal K}^{}_{E1M1}\delta_{ab}}{\sqrt{2\omega_12\omega_2}}(\cos\theta C^{11}_{20}-\sin\theta C^{11}_{12})(f^{110}_{2011}+f^{201}_{2011})\nonumber\\
&&\times(-1)^{1+m^{}_f}\left(
                                                                \begin{array}{ccc}
                                                                  1 & 1 & 1 \\
                                                                  m^{}_i+m^{}_s & -m & -m^{}_f \\
                                                                \end{array}
                                                              \right)
(q^{}_1\omega^{}_2+q^{}_2\omega^{}_1)^{}_{m},
\label{SPA-E1M1-1D-1P}
\end{eqnarray}
where the matrices are the Wigner 3-j symbols.

Then, after certain calculations, we get the $M^{}_{\pi\pi}$ distributions
\begin{eqnarray}                             
\label{dGamma/dM_pipi_E1M1_2S1P'}
&&\frac{d\Gamma^{}_{E1M1}(2^3S_1\to 1^1P_1)_{SPA}}{dM^{}_{\pi\pi}}=\frac{|{\cal K}^{}_{E1M1}|^2}{540\pi^3m_c^2}\left|\cos\theta C^{11}_{10}(f^{110}_{2011}+f^{001}_{2011})\right|^2\frac{M^{}_{J/\psi}}{M^{}_{\psi'}}M^{}_{\pi\pi}{\cal F}^{}_0{\cal F}^{}_3,
\nonumber\\
&&\frac{d\Gamma^{}_{E1M1}(1^1P_1\to 1^3S_1)_{SPA}}{dM_{\pi\pi}}=\frac{|{\cal K}^{}_{E1M1}|^2}{540\pi^3m_c^2}\left|\sin\theta C^{11}_{10}(f^{010}_{1110}+f^{101}_{1110})\right|^2\frac{M^{}_{J/\psi}}{M^{}_{\psi'}}M^{}_{\pi\pi}{\cal F}^{}_0{\cal F}^{}_3,\nonumber\\
&&\frac{d\Gamma^{}_{E1M1}(1^3D_1\to 1^1P_1)_{SPA}}{dM_{\pi\pi}}=\frac{|{\cal K}^{}_{E1M1}|^2}{1080\pi^3m_c^2}\left|(\cos\theta C^{11}_{20}-\sin\theta C^{11}_{12})(f^{110}_{1211}+f^{201}_{1211})\right|^2\frac{M^{}_{J/\psi}}{M^{}_{\psi'}}M^{}_{\pi\pi}{\cal F}^{}_0{\cal F}^{}_3,
\end{eqnarray}
where
\bea                                        
\hspace{1cm}{\cal F}^{}_3=\left(M_{\pi\pi}^2-4m_{\pi}^2\right)^{2}\Big[4\frac{K_0^4}{M_{\pi\pi}^4}-3\frac{K^2_0}{M_{\pi\pi}^2}-1\Big]
+40\left(M_{\pi\pi}^2-m_{\pi}^2\right)m_{\pi}^2\frac{K_0^2}{M_{\pi\pi}^2}\left(\frac{K_0^2}{M_{\pi\pi}^2}-1\right).
\label{I_3}
\eea

Integrating (\ref{dGamma/dM_pipi_E1M1_2S1P'}) over $dM^{}_{\pi\pi}$, we obtain the transition rates. For example,
\begin{eqnarray}                             
\label{Gamma_E1M1_2S1P'}
&&~~\Gamma^{}_{E1M1}(2^3S_1\to 1^1P_1)_{SPA}=\frac{|{\cal K}^{}_{E1M1}|^2}{540\pi^3}\left|\cos\theta C^{11}_{10}(f^{110}_{2011}+f^{001}_{2011})\right|^2\frac{M^{}_{J/\psi}}{M^{}_{\psi'}}\int^{\Delta M}_{2m^{}\pi}M^{}_{\pi\pi}{\cal F}^{}_0{\cal F}^{}_3dM^{}_{\pi\pi}.
\end{eqnarray}

Now the problem is to determine the unknown constant ${\cal K}^{}_{E1M1}$. So far there is no accurate enough data to determine it, so we should take the help of the 2GA.

\null\noindent
{\bf ii. 2GA}

In the 2GA, the hadronization factor can be expressed as
\begin{eqnarray}                                 
\langle\pi\pi|E_{-m_2}B_{-m_1}|0\rangle\approx\langle gg|E_{-m_2}B_{-m_1}|0\rangle
=\frac{\omega\epsilon^{}_{-m_2}(\lambda)(\bm q'\times\bm\epsilon'(\lambda'))_{-m_1}
+\omega'\epsilon'^{}_{-m_2}(\lambda')(\bm q\times\bm\epsilon(\lambda))_{-m_1}}
{\sqrt{2\omega}\sqrt{2\omega'}}.
\label{H-E1M1}
\end{eqnarray}
After certain calculations, we obtain
\begin{eqnarray}                                
\hspace{-0.4cm}\Gamma^{}_{E1M1}(2^3S_1\to1^1P_1\pi\pi)^{}_{2GA}&=&\frac{1}{2}\frac{N^2-1}{(2\pi)^5}
\left(\frac{g^{}_Eg^{}_M}{4Nm_c}\right)^2\frac{128\pi^2}{9}
\left|\cos\theta C^{11}_{10}(f^{110}_{2011}+f^{001}_{2011})\right|^2
\left\{\int_{0}^{\Delta M}\omega^3(\Delta M-\omega)^3d
\omega\right\}\nonumber\\
&&\hspace{-0.4cm}=6\frac{(N^2-1)}{27\pi^3}\left(\frac{g^{}_Eg^{}_M}{4Nm_c}\right)^2\left|\cos\theta C^{11}_{10}(f^{110}_{2011}+f^{001}_{2011})\right|^2
\frac{(\Delta M)^7}{140}.
\label{Gamma(2S-1Ppipi)-2gluon}
\end{eqnarray}
Compared with Eq.\,(\ref{2GA_Gamma(psi'-psi)}), we have the ratio
\bea                                          
&&\hspace{-1.6cm}R^{2GA}_{E1M1}=\left.\frac{\Gamma^{}_{E1M1}(2^3S_1\to 1^1P_1\pi\pi)}{\Gamma^{}_{E1E1}(2^3S_1\to 1^3S_1\pi\pi)}\right|_{2GA}
=6\left(\frac{g^{}_M}{2m^{}_cg^{}_E}\right)^2
\frac{\left|\cos\theta C^{11}_{10}(f^{110}_{2011}+f^{001}_{2011})\right|^2}{\left|\cos^2\theta f^{111}_{2010}+\displaystyle\frac{2}{5}\sin^2\theta f^{111}_{1210}\right|^2}.
\label{2GA E1M1ratio}
\eea
From (\ref{Gamma_E1M1_2S1P'}) and (\ref{psi'rate}) we can get the corresponding ratio
\bea                                           
R^{SPA}_{E1M1}=\left.\frac{\Gamma^{}_{E1M1}(2^3S_1\to 1^1P_1\pi\pi)}{\Gamma^{}_{E1E1}(2^3S_1\to 1^3S_1\pi\pi)}\right|_{SPA}
=\frac{\displaystyle\frac{|{\cal K}^{}_{E1M1}|^2}{540\pi^3m_c^2}\left|\cos\theta C^{11}_{10}(f^{110}_{2011}+f^{001}_{2011})\right|^2\frac{M^{}_{J/\psi}}{M^{}_{\psi'}}\int^{\Delta M}_{2m^{}\pi}M^{}_{\pi\pi}{\cal F}^{}_0{\cal F}^{}_3dM^{}_{\pi\pi}}
{|{\cal A}|^2\bigg[\cos^2\theta ~G^{}_{\cal AB}(\psi^\prime)
~|f^{111}_{2010}(\psi^\prime)|^2
+\sin^2\theta ~H^{}_{\cal K}(\psi^{\prime})
~|f^{111}_{1210}(\psi^{\prime})|^2\bigg]}.
\label{SPA E1M1ratio}
\eea

As has been argued \cite{KY81,KTY88}, we expect $R^{2GA}_{E1M1}\approx R^{SPA}_{E1M1}$. Thus from (\ref{2GA E1M1ratio}) and (\ref{SPA E1M1ratio}) we have
\bea                                          
\left|{\cal K}^{}_{E1M1}\right|^2&=&
\frac{\displaystyle 6\left(\frac{g^{}_M}{2m^{}_cg^{}_E}\right)^2}{\left|\cos^2\theta f^{111}_{2010}+\displaystyle\frac{2}{5}\sin^2\theta f^{111}_{1210}\right|^2}\nonumber\\
&&\times\frac{\displaystyle |{\cal A}|^2\bigg[\cos^2\theta ~G^{}_{\cal AB}(\psi^\prime)
~|f^{111}_{2010}(\psi^\prime)|^2
+\sin^2\theta ~H^{}_{\cal C}(\psi^{\prime})
~|f^{111}_{1210}(\psi^{\prime})|^2\bigg]}{\displaystyle\frac{1}{540\pi^3}\frac{M^{}_{J/\psi}}{M^{}_{\psi'}}\int^{\Delta M}_{2m^{}\pi}M^{}_{\pi\pi}{\cal F}^{}_0{\cal F}^{}_3dM^{}_{\pi\pi}}.
\label{K_E1M1 expression}
\eea
So $\left|{\cal K}^{}_{E1M1}\right|^2$ is expressed in terms of ${\cal A},{\cal B}$ and ${\cal C}$ appearing in $G_{{\cal AB}}$ ahd $H_{\cal C}$ [cf. (\ref{G,H})] which will be determined in Sec. IV.
\end{widetext}

\begin{widetext}
\subsection{Transition Rates of the $\bm{O(\delta_c^1)}$ M1CEDM1 processes (\ref{2S-1SEDM})}
In this transition, $\delta^{}_c$ is from the CEDM1 vertex. The transition amplitude is
\begin{eqnarray}                      
{\cal M}^{}_{M1cEDM1}&=&\frac{g^{}_M\delta_c}{24m_c^2}\delta^{}_{ab}
\sum_{KLM_s}\frac{1}{M_{\psi'}-E_{KL}}\left\{\langle \Phi^{}_F|
(s^{}_c-s^{}_{\bar c})_{m_2}
|KLM^{}_s\rangle
\langle KLM^{}_s|
(s^{}_c-s^{}_{\bar c})_{m_1}|\Phi^{}_I\rangle\right.\nonumber\\
&&\left.+
\langle \Phi^{}_F|
(s^{}_c-s^{}_{\bar c})_{m_1}
KLM^{}_s\rangle\langle KLM^{}_s|
(s^{}_c-s^{}_{\bar c})_{m_2}|\Phi^{}_I\rangle\right\}\langle\pi\pi|E_{-{m_2}}B_{-m_1}|0\rangle\nonumber\\
&=&\frac{g_M\delta^{}_c}{6m^2_c}\delta^{}_{ab}f^{000}_{2010}
\left\{a^{}_{m_2}(m_f)a^\ast_{m_1}(m_i)+
a^{}_{m_1}(m_f)a^\ast_{m_2}(m_i)\right\}\langle\pi\pi|E_{-{m_2}}B_{-m_1}|0\rangle.
\label{E1CEDM1amplitude}
\end{eqnarray}
\end{widetext}
The last step is obtained after certain evaluations of the MGE factor.

\null\noindent
{\bf i. SPA}

In (\ref{E1CEDM1amplitude}), the hadronization factor is the same as in the E1M1 transition. If we take the lowest order SPA expression
(\ref{E1M1-SPA}), we see that it is antisymmetric in $m^{}_1$ and $m^{}_2$. However, the MGE factor in (\ref{E1CEDM1amplitude}) is now symmetric in
$m^{}_1$ and $m^{}_2$, so that the lowest order SPA does not contribute to (\ref{E1CEDM1amplitude}). In this case, we should consider the next term in the SPA for the hadronization factor in (\ref{E1CEDM1amplitude}), i.e., the $O(q^3)$ term. Considering that $\langle\pi\pi|E_{-{m_2}}B_{-m_1}|0\rangle$ is a
second rank pseudo tensor, we have
\begin{widetext}
\bea                               
&&\null\hspace{-0.5cm}\frac{g^{}_M}{24}\delta^{}_{ab}\langle\pi\pi|E_{-{m_2}}B_{-m_1}|0\rangle=i\frac{{\cal K}^{}_{M1CEDM1}}{\Delta M\sqrt{2\omega^{}_1 2\omega^{}_2}}
\bigg[(q^{}_1-q^{}_2)^{}_{-m_1}(\bm q_1\times \bm q^{}_2)^{}_{-m^{}_2}+(q^{}_1-q^{}_2)^{}_{-m_2}(\bm q_1\times \bm q^{}_2)^{}_{-m^{}_1}\bigg]\nonumber\\
&&\hspace{0cm}=i\frac{{\cal K}^{}_{M1CEDM1}}{\Delta M\sqrt{2\omega^{}_1 2\omega^{}_2}}\bigg[(q^{}_1-q^{}_2)^{}_{-m_1}\epsilon^{}_{-m^{}_2,-m^{}_3,-m^{}_4}q^{}_{1m^{}_3}q^{}_{2m^{}_4}+(q^{}_1-q^{}_2)^{}_{-m_2}
\epsilon^{}_{-m^{}_1,-m^{}_3,-m^{}_4}q_{1m^{}_3}q^{}_{2m^{}_4}\bigg].
\label{M1CEDM1-SPA}
\eea
\end{widetext}
Here we have put a factor $i$ reflecting $\bm\partial\sim i\bm q$ in the expansion for convenience, and have introduced a scale parameter $\Delta M\equiv M^{}_{\psi'}-M^{}_{J/\psi}=\max(\omega_1+\omega_2)$ for making ${\cal K}^{}_{M1CEDM1}$ dimensionless like other SPA coefficients.
This expression is symmetric in $m^{}_1$ and $m^{}_2$ respecting the Bose symmetry between the two pions. It gives nonvanishing contribution to
(\ref{E1CEDM1amplitude}).

With (\ref{M1CEDM1-SPA}), after certain calculations, we obtain
\begin{widetext}
\begin{eqnarray}                       
\label{dGamma/dM_pipi-M1CEDM1}
&&\hspace{-2cm}\frac{d\Gamma^{}_{M1CEDM1}(2^3S_1\to1^3S_1\pi\pi)^{}_{SPA}}{dM_{\pi\pi}}=
\frac{8\left|{\cal K}^{}_{M1CEDM1}\right|^2}{3\pi^3|\Delta M|^2}\frac{\delta_c^2}{m_c^4}\frac{M^{}_{J/\psi}}{M^{}_{\psi'}}|f_{2010}^{000}|^2{\cal F}^{}_4,\\
\label{F_4}
&&\hspace{-2cm}{\cal F}^{}_4\equiv M^{}_{\pi\pi}
\Bigg\{-4M_{\pi\pi}^2{\cal F}_2
+\bigg[K_0^2(M_{\pi\pi}^2+4m_\pi^2)+2M_{\pi\pi}^2(M_{\pi\pi}^2-4m_{\pi}^2)\bigg]{\cal F}_1\nonumber\\
&&\hspace{1.2cm}
-(K_0^2+M_{\pi\pi}^2-4m_{\pi}^2)\left[\frac{1}{4}M_{\pi\pi}^2(M_{\pi\pi}^2-4m_{\pi\pi}^2)+m_{\pi}^2K_0^2\right]{\cal F}_0\Bigg\}.
\end{eqnarray}

Integrating (\ref{dGamma/dM_pipi-M1CEDM1}) over $dM^{}_{\pi\pi}$, we obtain the transition rate
\begin{eqnarray}                       
\label{Gamma-M1CEDM1}
&&\Gamma^{}_{M1CEDM1}(2^3S_1\to1^3S_1\pi\pi)^{}_{SPA}=
\frac{8\left|{\cal K}^{}_{M1CEDM1}\right|^2}{3\pi^3|\Delta M|^2}\frac{\delta_c^2}{m_c^4}\frac{M^{}_{J/\psi}}{M^{}_{\psi'}}|f_{2010}^{000}|^2\int^{\Delta M}_{2m^{}_\pi}{\cal F}^{}_4d M^{}_{\pi\pi}.
\end{eqnarray}
\end{widetext}

To determine ${\cal K}^{}_{M1CEDM1}$, we do the calculation with the help of the 2GA.

\null\noindent
{\bf ii. 2GA}

The 2GA expression for $\langle\pi\pi|E_{-{m_2}}B_{-m_1}|0\rangle$ has already been given in (\ref{H-E1M1}). After certain calculations, we have
\begin{widetext}
\begin{eqnarray}                                
\hspace{-0.5cm}\Gamma^{}_{M1CEDM1}(2^3S_1\to1^3S_1\pi\pi)^{}_{2GA}&=&
\frac{32}{9\pi^3}\left(\frac{g_M\delta_c}{6m^2_c}\right)^2\left|f^{000}_{2010}\right|^2
\frac{(\Delta M)^7}{140}.
\label{Gamma_{M1EDM1}(2S-1Spipi)-2gluon}
\end{eqnarray}

Compared with Eq.\,(\ref{2GA_Gamma(psi'-psi)}), we get the ratio
\begin{eqnarray}                                 
R^{2GA}_{M1CEDM1}(2^3S_1\to1^3S_1\pi\pi)&=&\left.\frac{\Gamma^{}_{M1CEDM1}(2^3S_1\to1^3S_1\pi\pi)}
{\Gamma^{}_{E1E1}(2^3S_1\to1^3S_1\pi\pi)}\right|^{}_{2GA}\nonumber\\
&=&12\frac{g^2_M\delta_c^2}{g_E^4m_c^4}\frac{|f^{000}_{2010}|^2}{\left|\cos^2\theta f^{111}_{2010}+\displaystyle\frac{2}{5}\sin^2\theta f^{111}_{1210}\right|^2}.
\label{ratio}
\end{eqnarray}

Taking $R^{2GA}_{M1CEDM1}\approx R^{SPA}_{M1CEDM1}$, we determine
\bea                                          
\left|{\cal K}^{}_{M1CEDM1}\right|^2&=&12\frac{g^2_M}{g_E^4}\frac{|\Delta M|^2|f^{000}_{2010}|^2}{\left|\cos^2\theta f^{111}_{2010}+\displaystyle\frac{2}{5}\sin^2\theta f^{111}_{1210}\right|^2}\nonumber\\
&&\times\frac{\displaystyle |{\cal A}|^2\bigg[\cos^2\theta ~G^{}_{\cal AB}(\psi^\prime)
~|f^{111}_{2010}(\psi^\prime)|^2
+\sin^2\theta ~H^{}_{\cal K}(\psi^{\prime})
~|f^{111}_{1210}(\psi^{\prime})|^2\bigg]}{\displaystyle\frac{8}{3\pi^3}\frac{M^{}_{J/\psi}}{M^{}_{\psi'}}|f_{2010}^{000}|^2\int^{\Delta M}_{2m^{}_\pi}{\cal F}^{}_4d M^{}_{\pi\pi}},
\label{K_M1CEDM1 expression}
\eea
in which $\left|{\cal K}^{}_{M1CEDM1}\right|^2$ is expressed in terms of ${\cal A},{\cal B}$ and ${\cal C}$ appearing in $G_{{\cal AB}}$ ahd $H_{\cal C}$ [cf. (\ref{G,H})].

\subsection{Transition Rates of the $\bm{O(\delta_c^1)}$ E1CEDM2 processes (\ref{2S-1S E1-CEDM2}) and (\ref{1D-1S E1-CEDM2})}

The two processes (\ref{2S-1S E1-CEDM2}) and (\ref{1D-1S E1-CEDM2}) for $2^3S_1\to 1^3S_1\pi\pi$ and $1^3D_1\to 1^3S_1\pi\pi$ belong to the E1CEDM2 transitions (the definition of CEDM2 is similar to that of M2 in $\eta(\pi)$-transitions \cite{KY81,KYF,Kuang06,KTY88}). The transition amplitude is of the form
\begin{eqnarray}                                   
\hspace{-0.5cm}{\cal M}_{E1CEDM2}&=&\frac{g^{}_E\delta^{}_c}{12m_c^{}}
\delta_{ab}\sum_{KLM_s}\frac{1}{M_{\Psi'}-E_{KL}}
\left\{\langle\Phi^{}_F\pi\pi|\bm S\cdot\bm E(\bm r\cdot
\bm\partial)|KLM_s\rangle
\langle KLM_s|\bm r'\cdot\bm E|\Phi^{}_I\rangle\right.\nonumber\\
&&+\left.\langle\Phi^{}_F\pi\pi|\bm r\cdot\bm E|\phi^{}_{KLM_s}\rangle
\langle KLM_s|\bm S\cdot\bm E(\bm r'\cdot
\bm\partial)|\Phi^{}_I\rangle\right\}\nonumber\\
&=&\frac{g^{}_E\delta^{}_c}{4Nm_c^{}}\delta_{ab}
\sum_{KLM_s}\frac{1}{M_{\Psi'}-E_{KL}}
\left\{\langle\Phi^{}_F|S_{m_1}r_{m_3}|KLMM_s\rangle
\langle KLM_s|r'_{m_2}|\Phi^{}_I\rangle
\langle\pi\pi|(\partial_{-m_3}E_{-m_1})E_{-m_2}|0\rangle\right.\nonumber\\
&&+\left.\langle\Phi^{}_F|r_{m_1}|KLM_s\rangle
\langle KLM_s|S_{m_2}r'_{m_3}|\Phi^{}_I\rangle\langle\pi\pi|E_{-m_1}\partial_{-m_3} E_{-m_2}|0\rangle\right\}.
\label{M_{E1CEDM2}}
\end{eqnarray}
\end{widetext}
where $\bm S$ is the total spin operator of $c$ and $\bar c$. The matrix element of $\bm S$ between two
quarkonium spin states can be evaluated by using Eq.~(A20) in the APPENDIX.

As before, we express the hadronzation factor in SPA and 2GA, respectively.

\null\noindent
{\bf i. SPA}

In this kind of transition, the hadronization factor is a third rank tensor. Considering the Bose symmetry between the two pions, its general form in the SPA is
\begin{widetext}
\begin{eqnarray}                              
\hspace{0cm}\frac{g^{}_E}{12}\langle\pi\pi| E_{-m_1}\partial_{-m_3}E_{-m_2}|0\rangle
&=&\frac{g^{}_E}{12}\langle\pi\pi| (\partial_{-m_3}E_{-m_1})E_{-m_2}|0\rangle\nonumber\\
&=&i\frac{{\cal K}^{}_{E1CEDM2}}{\sqrt{2\omega^{}_1 2\omega^{}_2}}\left(q^{}_{1,-m^{}_2}q^{}_{2,-m^{}_1}+q^{}_{1,-m^{}_1}q^{}_{2,-m^{}_2}\right)\left(q^{}_{1,-m^{}_3}+q^{}_{2,-m^{}_3}\right).
\label{H-E1EDM2}
\end{eqnarray}
Here we also put a factor $i$ reflecting $\bm\partial\sim i\bm q$ for convenience.

After certain evaluations, we obtain
\begin{eqnarray}                            
\label{dGamma/dM_pipi_E1CEDM2_2s1s}
\frac{d\Gamma^{}_{E1CEDM2}(2^3S_1\to 1^3S_1\pi\pi)^{}_{SPA}}{dM_{\pi\pi}}=\frac{8\left|{\cal K}^{}_{E1CEDM2}\right|^2}{27\pi^3}
\frac{\delta_c^2}{m_c^2} \frac{M^{}_{J/\psi}}{M^{}_{\psi'}}\left|f_{2010}^{111}\right|^2\Big[{\cal F}^{}_5+2{\cal F}^{}_6\Big]
\end{eqnarray}

\begin{eqnarray}                             
\label{dGamma/dM_pipi_E1CEDM2_1d1s}
&&\null\hspace{-0.4cm}\frac{d\Gamma^{}_{E1CEDM2}(1^3D_1\to 1^3S_1\pi\pi)^{}_{SPA}}{dM^{}_{\pi\pi}}
=\frac{8\left|{\cal K}^{}_{E1CEDM2}\right|^2}{2700\pi^3}\frac{\delta_c^2}{m_c^2}\frac{M^{}_{J/\psi}}{M^{}_{\psi'}}\left|f^{111}_{1210}\right|^2
\Bigg\{18K^2({\cal F}^{}_7+{\cal F}^{}_8)-13{\cal F}^{}_5+22{\cal F}^{}_6\Bigg\},
\end{eqnarray}
where
\begin{eqnarray}                            
&&\hspace{-0.6cm}{\cal F}^{}_5=
M^{}_{\pi\pi}\left\{2K_0^2{\cal F}_2-\left[\frac{M_{\pi\pi}^4}{2}+M_{\pi\pi}^2K_0^2-2m_\pi^2K_0^2\right]{\cal F}_1
+\Big[\frac{1}{4}M_{\pi\pi}^2(M_{\pi\pi}^2+4m_\pi^2)K_0^2-m_{\pi}^2K_0^4-\frac{M_{\pi\pi}^4}{2}m_\pi^2\Big]{\cal F}_0\right\},\nonumber\\
&&\hspace{-0.6cm}{\cal F}^{}_6=
M^{}_{\pi\pi}\left\{K_0^2{\cal F}_2-\left[(M_{\pi\pi}^2-m_{\pi}^2)K_0^2-\frac{M_{\pi\pi}^4}{4}\right]{\cal F}_1
+\frac{1}{8}M_{\pi\pi}^2(M_{\pi\pi}^2-2m_\pi^2)(2K_0^2-M_{\pi\pi}^2){\cal F}_0\right\}\nonumber\\
&&\hspace{-0.6cm}{\cal F}^{}_7=
M^{}_{\pi\pi}\left\{{\cal F}_2-(M_{\pi\pi}^2-2m_\pi^2){\cal F}_1+\frac{1}{4}(M_{\pi\pi}^2-2m_\pi^2)^{2}{\cal F}_0\right\}\nonumber\\
&&\hspace{-0.6cm}{\cal F}^{}_8=
M^{}_{\pi\pi}\left\{{\cal F}_2+2m_{\pi}^2{\cal F}_1+m_\pi^2(m_\pi^2-K_0^2){\cal F}_0\right\}.
  \label{I_5,I_6,I_7,I_8}
  \end{eqnarray}
 \end{widetext}

 Integrating (\ref{dGamma/dM_pipi_E1CEDM2_2s1s}) over $dM^{}_{\pi\pi}$, we obtain the transition rate of $\psi(2^3S_1)\to \psi(1^3S_1)\pi\pi$.
 \begin{eqnarray}                            
\label{Gamma-E1CEDM2_2s1s}
&&\Gamma^{}_{E1CEDM2}(2^3S_1\to 1^3S_1\pi\pi)^{}_{SPA}=\frac{8\left|{\cal K}^{}_{E1CEDM2}\right|^2}{27\pi^3}
\frac{\delta_c^2}{m_c^2}\nonumber\\
&&~~~~~~\times \frac{M^{}_{J/\psi}}{M^{}_{\psi'}}\left|f_{2010}^{111}\right|^2 \int_{2m^{}_\pi}^{\Delta M} \Big[{\cal F}^{}_5+2{\cal F}^{}_6\Big]dM^{}_{\pi\pi}.
\end{eqnarray}

Next, we determine the unknown constant ${\cal K}^{}_{E1CEDM2}$ with the help of the 2GA.

\null\noindent
{\bf ii. 2GA}

In the 2GA
\begin{widetext}
\bea                                
&&\langle \pi\pi|E^a_{-m^{}_1}\partial^{}_{-m^{}_3}E^b_{-m^{}_2}|0\rangle=\frac{-i\delta^{}_{ab}\omega^{}_1\omega^{}_2}{\sqrt{2\omega^{}_12\omega^{}_2}}
\bigg[\epsilon^{}_{1,-m^{}_2}(\lambda^{}_1)\epsilon^{}_{2,m^{}_1}(\lambda^{}_2)q^{}_{2,-m^{}_3}
+\epsilon^{}_{2,-m^{}_2}(\lambda^{}_2)\epsilon^{}_{1,-m^{}_1}(\lambda^{}_1)q^{}_{1,-m^{}_3}\bigg],\nonumber\\
&&\langle \pi\pi|(\partial^{}_{m^{}_3}E^a_{-m^{}_1})E^b_{-m^{}_2}|0\rangle=\frac{-i\delta^{}_{ab}\omega^{}_1\omega^{}_2}{\sqrt{2\omega^{}_12\omega^{}_2}}
\bigg[\epsilon^{}_{1,-m^{}_2}(\lambda^{}_1)\epsilon^{}_{2,m^{}_1}(\lambda^{}_2)q^{}_{1,-m^{}_3}
+\epsilon^{}_{2,-m^{}_2}(\lambda^{}_2)\epsilon^{}_{1,-m^{}_1}(\lambda^{}_1)q^{}_{2,-m^{}_3}\bigg].
\label{2GA-E1CEDM2}
\eea
With this, we obtain
\bea                                
&&\Gamma^{}_{E1CEDM2}(2^3S_1\to 1^3S_1\pi\pi)^{}_{2GA}=\left(\frac{g^{}_E}{3}\right)^2
\frac{\delta_c^2}{m_c^2}\left|f^{111}_{2010}\right|^2\frac{(\Delta M)^9}{5103\pi^3}.
\label{Gamma_E1CEDM2 2S1S-2GA}
\eea
Compared with Eq.\,(\ref{2GA_Gamma(psi'-psi)}), we have the ratio
\begin{eqnarray}                                 
R^{2GA}_{E1CEDM2}(2^3S_1\to1^3S_1\pi\pi)^{}_{2GA}&=&\left.\frac{\Gamma^{}_{E1CEDM2}(2^3S_1\to1^3S_1\pi\pi)}
{\Gamma^{}_{E1E1}(2^3S_1\to1^3S_1\pi\pi)}\right|^{}_{2GA}\nonumber\\
&=&\frac{10}{27}\left(\frac{\Delta M}{g^{}_E}\right)^2\frac{\delta_c^2}{m_c^2}\frac{|f^{111}_{2010}|^2}{\left|\cos^2\theta f^{111}_{2010}+\displaystyle\frac{2}{5}\sin^2\theta f^{111}_{1210}\right|^2}.
\label{ratio}
\end{eqnarray}
Taking $R^{2GA}_{E1CEDM2}\approx R^{SPA}_{E1CEDM2}$, we have
\bea                                          
\left|{\cal K}^{}_{E1CEDM2}\right|^2&=&\frac{5}{4}\left(\frac{\Delta M}{g^{}_E}\right)^2\frac{|f^{111}_{2010}|^2}{\left|\cos^2\theta f^{111}_{2010}+\displaystyle\frac{2}{5}\sin^2\theta f^{111}_{1210}\right|^2}\nonumber\\
&&\times\frac{\displaystyle |{\cal A}|^2\bigg[\cos^2\theta ~G^{}_{\cal AB}(\psi^\prime)
~|f^{111}_{2010}(\psi^\prime)|^2
+\sin^2\theta ~H^{}_{\cal K}(\psi^{\prime})
~|f^{111}_{1210}(\psi^{\prime})|^2\bigg]}{\displaystyle\frac{1}{\pi^3}
\frac{M^{}_{J/\psi}}{M^{}_{\psi'}}\left|f_{2010}^{111}\right|^2 \int_{2m^{}_\pi}^{\Delta M} \Big[{\cal F}^{}_5+2{\cal F}^{}_6\Big]dM^{}_{\pi\pi}}.
\label{K_E1CEDM2 expression}
\eea
\end{widetext}
So ${\cal K}^{}_{E1CEDM2}$ is expressed in terms of ${\cal A},{\cal B}$ and ${\cal C}$ appearing in $G_{{\cal AB}}$ ahd $H_{\cal C}$ [cf. (\ref{G,H})].

Now, in the SPA, we have expressed all the unknown constants $\left|{\cal K}^{}_{E1M1}\right|^2, \left|{\cal K}^{}_{M1CEDM2}\right|^2$ and $\left|{\cal K}^{}_{E1CEDM2}\right|^2$ occurring in the $O(\delta_c^1)$ transitions in terms of ${\cal A}, {\cal B}$ and ${\cal C}$ in the $O(\delta_c^0)$ transitions. The values of ${\cal K}^{}_{E1M1}, {\cal K}^{}_{M1CEDM2}$ and ${\cal K}^{}_{E1CEDM2}$ from the values of ${\cal A}, {\cal B}$ and ${\cal C}$ obtained from the best fit of the $O(\delta_c^0)$ contribution to the experimental data are listed in TABLE\,\ref{Kvalues} in APPENDIX B.  Therefore, the magnitudes of the $O(\delta_c^1)$ transitions are characterized by only one parameter $\delta^{}_c$.

\section{Determination of $\bm{\delta_c}$ from the BESII Data of $\bm{\psi'\to J/\psi+\pi^++\pi^-}$}

In Sec.\,III, we calculated the transition rates contributed by the mechanisms (\ref{2S-1Spipi})$\--$(\ref{1D-1S E1-CEDM2}) individually, and expressed all the unknown coefficients in the SPA in terms of the parameters ${\cal A},\,{\cal B},$ and ${\cal C}$. Now we are going to determine the parameters ${\cal A},\,{\cal B},\,{\cal C},$ and $\delta_c$ from the best fit of the theoretical prediction to the experimental data.

The transition amplitudes are functions of the two pion momenta $\bf q_1$ and $\bf q_2$. For two transition amplitudes, if their pion momenta dependence belong to the same representation of the spacial rotation and reflection symmetries, they may have nonvanishing interference term in the $M_{\pi\pi}$ distribution. So that we should take account of such interference terms in calculating the $M_{\pi\pi}$ distribution. Specifically, there are three kinds of interference terms to be considered, namely {\it (i) E1M1-CEDM2 interference, (ii) CEDM2-M1CEDM1 interference, and (iii) interference between the $2^3S_1\to 1^3S_1\,\pi\pi$ and $1^3D_1\to 1^3S_1\,\pi\pi$ amplitudes in E1CEDM2}. So we should consider the following total transition amplitude:

\begin{widetext}
\begin{eqnarray}                                
\label{Mtotal}
\null\hspace{-1cm}{\cal M}^{}_{tot}(\psi'\to J/\psi\,\pi\pi)&=&\Bigg[\cos\theta C^{20}_{20}C^{10}_{10}{\cal M}_{E1E1}(2^3S_1\to 1^3S_1)-\sin\theta C^{12}_{12}C^{10}_{10}{\cal M}_{E1E1}(1^3D_1\to 1^3S_1)\Bigg]\nonumber\\
&&+\Bigg[\cos\theta C^{20}_{20}C^{10}_{10}{\cal M}_{M1CEDM1}(2^3S_1\to 1^3S_1)+\cos\theta C^{20}_{20}C^{10}_{10}{\cal M}_{E1CEDM2}(2^3S_1\to 1^3S_1)\nonumber\\
&&-\sin\theta C^{12}_{12}C^{10}_{10}{\cal M}_{E1CEDM2}(1^3D_1\to 1^3S_1)
+\cos\theta
C_{10}^{11}C^{20}_{20}{\cal M}_{E1M1}(2^3S_1\to 1^1P_1)\nonumber\\
&&+\Big(\cos\theta
C_{20}^{11}-\sin\theta
C^{11}_{12}\Big)C^{10}_{10}{\cal M}_{E1M1}(1^1P_1\to 1^3S_1)-\sin\theta
C^{11}_{10}C^{12}_{12}{\cal M}_{E1M1}(1^3D_a\to 1^1P_1)\Bigg].
\end{eqnarray}
\end{widetext}
We see from Eqs.\,(\ref{dGamma/dM_pipi}),\,(\ref{dGamma/dM_pipi_E1M1_2S1P'}),\,(\ref{dGamma/dM_pipi-M1CEDM1}),\,(\ref{dGamma/dM_pipi_E1CEDM2_2s1s}), and (\ref{dGamma/dM_pipi_E1CEDM2_1d1s}) that the CEDM contributions to $d\Gamma(\psi'\to J/\psi\,\pi\pi)/dM^{}_{\pi\pi}$ are of $O(\delta_c^2)$. So that the mixing coefficients $C^{10}_{10},\,C^{20}_{20}$, and $C^{12}_{12}$ in the first square bracket on the right hand side of Eq.\,(\ref{Mtotal}) should be expanded up to their $O(\delta_c^2)$ terms which give the CEDM contributions through the normalization of the mixing coefficients, i.e., we should take
\begin{widetext}
\begin{eqnarray}                              
&&\displaystyle C_{10}^{10}=\frac{1}{\sqrt{1+\displaystyle\left(\frac{_0\langle 1^1P_1|V_1
|1^3S_1\rangle_0}
{E^0_{1^1P_1}-E^0_{1^3S_1}}\right)^2}}=1-\frac{1}{2}\left|C^{11}_{10}\right|^2+O(\delta_c^4),\nonumber\\
&&\displaystyle C_{20}^{20}=\frac{1}{\sqrt{1+\displaystyle\left(\frac{_0\langle 1^1P_1|V_1
|2^3S_1\rangle_0}
{E^0_{2^3S_1}-E^0_{1^1P_1}}\right)^2}}=1-\frac{1}{2}\left|C^{11}_{20}\right|^2+O(\delta_c^4),\nonumber\\
&&\displaystyle C_{12}^{12}
=\frac{1}{\sqrt{1+\displaystyle\left(\frac{_0\langle 1^1P_1|V_1|1^3D_1\rangle_0}
{E^0_{1^3D_1}-E^0_{1^1P_1}}\right)^2}}=1-\frac{1}{2}\left|C^{11}_{12}\right|^2+O(\delta_c^4),
\label{Cexpansion}
\end{eqnarray}
\end{widetext}
in the first square bracket on the right hand side of (\ref{Mtotal}).

The contributions to $\displaystyle\frac{d\Gamma(\psi'\to J/\psi\,\pi\pi)}{dM^{}_{\pi\pi}}$ from individual mechanisms have been given in Eqs.\,(\ref{dGamma/dM_pipi}), (\ref{dGamma/dM_pipi_E1M1_2S1P'}),\,(\ref{dGamma/dM_pipi-M1CEDM1}),\,(\ref{dGamma/dM_pipi_E1CEDM2_2s1s}), and (\ref{dGamma/dM_pipi_E1CEDM2_1d1s}). Now we consider the contributions from the interference terms. By the same approach as in Sec.\,III, we obtain
\begin{widetext}
\null\noindent{\bf (i) E1M1-E1CEDM2:}
\begin{eqnarray}                           
\null\hspace{-0.6cm}\left.\frac{d\Gamma^{}_{E1M1-E1CEDM2}}{dM_{\pi\pi}}\right|_{SPA}&=&{\cal K}_{E1M1}{\cal K}_{E1CEDM2}S^{}_{E1M1}\left\{\displaystyle\frac{2\sqrt{6}M_{J/\psi}M_{\pi\pi}}{27\pi^3m^{2}_cM_{\psi'}}cos\theta f^{111}_{2010}{\cal F}_9\right.\nonumber\\
&&\left.+\frac{2\sqrt{3}M^{}_{J/\psi}M^{}_{\pi\pi}}{45\pi^3m^{2}_cM^{}_{\psi'}}\cos\theta f^{111}_{2010}{\cal F}_{10}\right\},
\label{Gamma_E1M1-E1CEDM2}
\end{eqnarray}
where
\begin{eqnarray}                           
&&S^{}_{E1M1}\equiv \sqrt{2}\cos\theta C^{11}_{10}(f^{110}_{2011}+f^{001}_{2011})-\sqrt{2}(\cos\theta C^{11}_{20}-\sin\theta C^{11}_{12})(f^{101}_{1110}+f^{010}_{1110})-\sin\theta C^{11}_{10}(f^{110}_{1211}+f^{201}_{1211}),\label{S(E1M1)}\\
&&{\cal F}_9\equiv K_0[4{\cal F}_2-(2M_{\pi\pi}^2-4m_{\pi}^2){\cal F}_1+(\frac{1}{4}M_{\pi\pi}^4-m_{\pi}^2K_0^2){\cal F}_0] ,
\label{F_9}\\
&&{\cal F}_{10}\equiv \frac{1}{6}K_0[16{\cal F}_2-(11M_{\pi\pi}^2-23m_{\pi}^2){\cal F}_1+(\frac{7}{4}M_{\pi\pi}^4-3m_{\pi}^2M_{\pi\pi}^2-m_{\pi}^2K_0^2){\cal F}_0].
\label{F_10}
\end{eqnarray}

\null\noindent{\bf (ii) M1CEDM1-E1CEDM2:}
\begin{eqnarray}                          
\left.\frac{d\Gamma^{}_{M1CEDM1-E1CEDM2}}{dM_{\pi\pi}}\right|_{SPA}=\frac{2\sqrt{2}M^{}_{J/\psi}M^{}_{\pi\pi}}{5\pi^3M^{}_{\psi'}\Delta M}\frac{\delta_c^2}{m_c^3}{\cal K}^{}_{M1CEDM1}{\cal K}^{}_{E1CEDM2}\cos\theta \sin\theta f^{000}_{2010}f^{111}_{1210}{\cal F}_{11},
\label{Gamma_M1CEDM1-E1CEDM2}
\end{eqnarray}
where
\begin{eqnarray}                              
{\cal F}^{}_{11}&\equiv& \frac{1}{3}\{-4M^2_{\pi\pi}{\cal F}^{}_2
+\left[K^2_0(M^2_{\pi\pi}+4m_{\pi}^2)+2M^2_{\pi\pi}(M_{\pi\pi}^2-4m_\pi^2)
\right]{\cal F}^{}_1\nonumber\\
&&-(K_0^2+M_{\pi\pi}^2-4m_\pi^2)\left[\frac{1}{4}M_{\pi\pi}^2(M_{\pi\pi}^2-4m_\pi^2)+m_\pi^2K_0^2\right]{\cal F}^{}_0\}.
\label{F_11}
\end{eqnarray}

\null\noindent{\bf (iii) E1CEDM2-E1CEDM2:}
\begin{eqnarray}                              
\left.\frac{d\Gamma^{}_{E1CEDM2-E1CEDM2}}{dM_{\pi\pi}}\right|_{SPA}=-\frac{8\sqrt{2}M^{}_{J/\psi}M_{\pi\pi}}{45\pi^3M^{}_{\psi'}}\frac{\delta_c^2}{m_c^2}\left|{\cal K}^{}_{E1CEDM2}\right|^2\cos\theta \sin{\theta}f^{111}_{2010}f^{111}_{1210}{\cal F}^{}_{12},
\label{Gamma_E1CEDM2-E1CEDM2}
\end{eqnarray}
where
\begin{eqnarray}                              
{\cal F}^{}_{12}\equiv \frac{1}{3}\{-4M^2_{\pi\pi}{\cal F}^{}_2
+(K^2_0M^2_{\pi\pi}+2M^4_{\pi\pi}
+4m^2_{\pi}K_0^2-8m^2_{\pi}M^2_{\pi\pi}){\cal F}^{}_1-(M^2_{\pi\pi}-4m_{\pi}^2+K_0^2)(\frac{M^4_{\pi\pi}}{4}+m^2|K|^2){\cal F}^{}_0\}.
\label{F_12}
\end{eqnarray}
\end{widetext}

With all these results, we are ready to determine the unknown parameters by the best fit of the theoretical prediction to the experimental data.
We take the BESII data on $\psi'\to J/\psi\,\pi^+\,\pi^-$ \cite{BESII07} based on $1.4\times 10^7$ $\psi'$ events. We proceed the determination by taking the CEDM effect as a perturbation. We first take the 0th order $O(\delta_c^0)$ contribution to fit the BESII data. There are three unknown parameters ${\cal A},\,{\cal B}$ and ${\cal C}$ in the 0th order contribution in which ${\cal A}$ is an overall normalization factor irrelevant to the $M^{}_{\pi\pi}$ distribution. It is determined by fitting the total transition rate with the experimental value \cite{PDG}.
 The ratios ${\cal B/A}$ and ${\cal C/A}$ do affect the $M^{}_{\pi\pi}$ distribution, and they are determined by the best fit of the theoretical $M^{}_{\pi\pi}$ distribution with the BESII data. The best fit curve in the CK model, as an example, is shown by the red dashed line in FIG.\,\ref{BestFit} together with the BESII data. 

\begin{figure}[h]
\includegraphics[width=8.5truecm,clip=true]{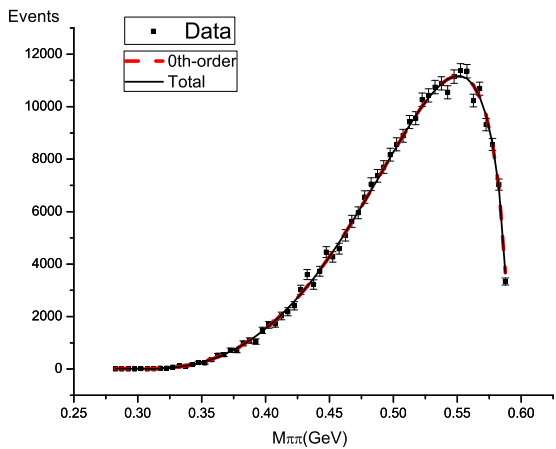}
\null\vspace{-0.5cm}
\caption{\footnotesize The best fit theoretical curves of the 0th order contribution (red dashed line) and the total contribution (dark solid line) in the CK model together with the BESII data on $d\Gamma(\psi'\to J/\psi\,\pi\pi)/dM^{}_{\pi\pi}$ \cite{BESII07}.}
\label{BestFit}
\end{figure}
The values of $|{\cal A}|$ determined from (\ref{GammaExpt}) and the values of ${\cal B}/{\cal A}$ and ${\cal C}/{\cal A}$ determined from the best fit of the $O(\delta_c^0)$ contribution to the $M^{}_{\pi\pi}$ distribution are:
\begin{eqnarray}                       
&&\null\hspace{-0.8cm}{\rm CK~ Model}:~~~~~~|{\cal A}|=1.641 \pm 0.036\nonumber\\
&&\null\hspace{-0.2cm}{\cal B/A}=-0.372\pm0.006,~~|{\cal C/A}|=0.855\pm0.020,\nonumber\\
&&\null\hspace{-0.8cm}{\rm Cornell~ model}:~|{\cal A}|=2.09 \pm 0.05\nonumber\\
&&\null\hspace{-0.2cm}{\cal B/A}=-0.371\pm0.006,~~
|{\cal C/A}|=1.000\pm0.166.
\label{B/A,C/A}
\end{eqnarray}
The values of the SPA coefficients ${\cal K}^{}_{E1M1}$. ${\cal K}^{}_{M1CEDM1}$, and ${\cal K}^{}_{E1CEDM2}$ obtained from the best fit values (\ref{B/A,C/A}) are listed in TABLE\,\ref{Kvalues} in APPENDIX\,B.

As an example, we plot the contributions from various $O(\delta_c^2)$ CEDM terms to the $M^{}_{\pi\pi}$ distribution in the CK model in FIG.\,\ref{CEDMs} with $\delta_c=1$ (the $\delta_c$-independent part). We see that the CEDM contribution increases the low $M^{}_{\pi\pi}$ distribution and reduces the high $M^{}_{\pi\pi}$ distribution, which is just the opposite to the 0th order distribution. This is the reason why the $M^{}_{\pi\pi}$ distribution of $\psi'\to J/\psi\,\pi^+\pi^-$ can sensitively determine the CEDM parameter $\delta_c$.

\begin{figure}[h]
\includegraphics[width=9truecm,clip=true]{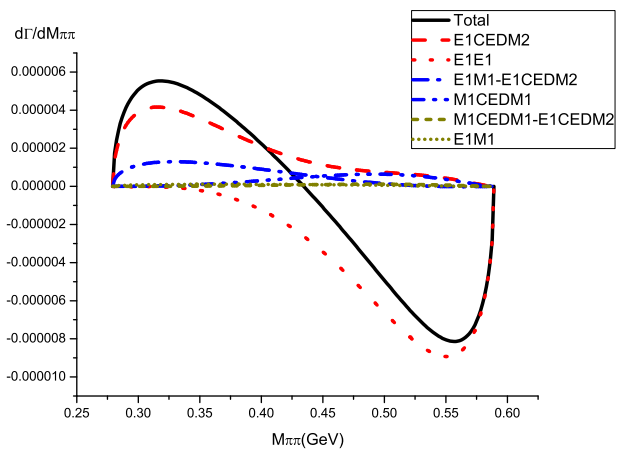}
\null\vspace{-0.5cm}
\caption{\footnotesize Various CEDM contributions to the $M_{\pi\pi}$ distribution for a given $\delta_c$ in the CK model.}
\label{CEDMs}
\end{figure}

Next we take into account the CEDM contributions Eqs.\,(\ref{dGamma/dM_pipi}), (\ref{dGamma/dM_pipi_E1M1_2S1P'}),\,(\ref{dGamma/dM_pipi-M1CEDM1}),\,(\ref{dGamma/dM_pipi_E1CEDM2_2s1s}), (\ref{dGamma/dM_pipi_E1CEDM2_1d1s}),(\ref{Cexpansion}),(\ref{Gamma_E1M1-E1CEDM2}), (\ref{Gamma_M1CEDM1-E1CEDM2}), and (\ref{Gamma_E1CEDM2-E1CEDM2}) to make the best fit of the total contribution. The best fit curve of the total contribution is shown by the dark solid line in FIG.\,\ref{BestFit}. The numerical result shows that the dark solid curve improves the fit a little bit (with slightly reduced $\chi^2$ value) although the difference between the two curves is too small to be visible in FIG.\,\ref{BestFit}. The best fit dark solid curve determines the best fit value of $\delta_c$ ($d'_c$) listed in TABLE\,\ref{delta_cvalue} in which the error bars of $\delta_c$ ($d'_c$) are determined from the experimental error bars in the $M^{}_{\pi\pi}$ distribution. Note that the best fit value of $\delta_c$ is nonvanishing. However, considering the error bars in TABLE\,\ref{delta_cvalue}, the obtained $\delta_c$ is still consistent with zero. So the model dependence of the present approach is not serious.\\\\

\begin{widetext}

\begin{table}[h]
\caption{The Best fit values of $\delta_c$ ($d'_c$) in the CK model and the Cornell model.}
\tabcolsep 1pt
\begin{tabular}{cccccccc}
\hline\hline
&&CK model&&&&Cornell model&\\
&68$\%$ C.L.&&95$\%$ C.L.&&
68$\%$ C.L.&&95$\%$ C.L.\\
\hline
$\left|\delta_c\right|$~~~~~~~~~~&0.025$\pm$0.295&&0.025$\pm$0.420&&
0.078$\pm$0.373&&0.078$\pm$0.544\\
$\left|d'_c\right|$ (e$\cdot$cm)&~(0.110$\pm$1.300)$\times 10^{-14}$&&(0.110$\pm$1.851)$\times 10^{-14}$&&~(0.276$\pm$1.320)$\times 10^{-14}$&&(0.276$\pm$1.926)$\times 10^{-14}$\\
\hline\hline
\label{delta_cvalue}
\end{tabular}
\end{table}
\end{widetext}


For instance, the $95\%$ C.L. upper bound of $d'_c$ is
\begin{eqnarray}                                         
&&{\rm CK~model}:~~~~~~~~\left|d'_c\right|<1.96\times 10^{-14}\,{\rm e\cdot cm}\nonumber\\
&&{\rm Cornell~model}:~~~\left|d'_c\right|<2.20\times 10^{-14}\,{\rm e\cdot cm}.
\label{UpperBound}
\end{eqnarray}
So the model dependence of the present approach is roughly $12\%$. 

We would like to mention that, in Eqs.\,(\ref{K_E1M1 expression}), (\ref{K_M1CEDM1 expression}), and (\ref{K_E1CEDM2 expression}), only the absolute values of the SPA coefficients ${\cal K}^{}_{E1M1},\,{\cal K}^{}_{M1CEDM1}$ and ${\cal K}^{}_{E1CEDM2}$ are determined. So that there is still an uncertain sign in (\ref{Gamma_E1M1-E1CEDM2}) and (\ref{Gamma_M1CEDM1-E1CEDM2}). Actually, If we were able to calculate the hadronization matrix elements from the first principles of QCD, there would not be such sign uncertainties. The present sign uncertainties are due to the phenomenological approach to the haronization factors taken in this paper as lacking of reliable QCD evaluation of the hadronization matrix elements.
 In TABLE\,\ref{delta_cvalue}, we only take the simple case that all the SPA coefficients are of the same sign. Now we consider how will the final result affected if they have different signs. First we see from FIG.\,\ref{CEDMs} that the contributions of (\ref{Gamma_M1CEDM1-E1CEDM2}) is so small that its uncertain sign only causes negligible effect in the total $M^{}_{\pi\pi}$ distribution. Thus only the uncertain sign in (\ref{Gamma_E1M1-E1CEDM2}) matters. If we take ${\cal K}_{E1M1}{\cal K}_{E1CEDM2}<0$
in (\ref{Gamma_E1M1-E1CEDM2}), the total CEDM contribution to $M^{}_{\pi\pi}$ distribution will be reduced, and thus the determined $\left|\delta_c\right|\,(\left|d'_c\right|)$ will be larger. Fortunately, we see from FIG.\,\ref{CEDMs} that the E1M1-CEDM1 contribution is smaller than the individual CEDM contributions. Our calculation shows that the determined $\left|\delta_c\right|$ at $68\%$ C.L. will change to $0.047\pm0.383$ for the CK model and $0.123\pm0.511$ for the Cornell model. Then the upper bound of $\left|d'_c\right|$ will change to $2.63\times 10^{-14}$ e\,cm for the CK models and $3.09\times 10^{-14}$ e\,cm, i.e., the uncertainty of the upper bound is $34\%$ and $40\%$ for the CK model and the Cornell model, respectively. Therefore the effect of the uncertain signs is not so serious. We conclude that {\it the $95\%$ C.L. upper bound of $\left|d'_c\right|$ determined from the BESII data is}
\begin{eqnarray}                     
\left| d'_c\right|<3\times 10^{-14} {\rm e\,cm}.
 \label{FinalBound}
 \end{eqnarray}
 {\it This is the first experimentally determined upper bound of the CEDM of the c quark.}

The BES detector has already been updated to BESIII with the efficiency of measuring low momentum pions significantly improved relative to BESII. So far BESIII has accumulated $1.06\times 10^8$ $\psi'$ events, and will be able to accumulate $7\times 10^8$ $\psi'$ events in the summer of 2012. That will be a huge sample. We expect that the new BESIII data may determine $\delta_c$ to a higher precision.


\section{The CP Odd Operator $\bm{{\cal O}}$}
We can propose another way of determining $\delta_c$ ($d'_c$) linearly from the data of  $\psi'\to J/\psi\,\pi\pi$. Consider the process
\begin{eqnarray}                        
e^+e^-\to\psi'\to J/\psi\,\pi^+\pi^-.
\label{FullProcess}
\end{eqnarray}
Let $\hat{\bm p} (-\hat{\bm p}),\,\hat{\bm q}_1$ and $\hat{\bm q}_2$ be the unit vectors of the momenta of the positron (electron), $\pi^+$ and $\pi^-$, respectively. For unpolarized $e^+$ and $e^-$ in the overall c.m. system, the initial state is then $\--$ in the sense of the density matrix $\--$ CP-even. Therefore any nonzero expectation value of a CP-odd correlation of the final state particles is an unambiguous indication of CP violation. With our assumption of the CEDM of the c quark, the expectation values of the CP-odd operators will be linear in $\delta_c(d'_c)$. On the other hand, to the expectation values of CP-even operators, the CEDM can only contribute in $O(\delta_c^2)$ or higher even powers.
We shall now construct a CP-odd operator for the reaction (\ref{FullProcess}) following Eqs. (3.20) in Ref.\,\cite{Nachtmann}
\begin{eqnarray}                         
{\cal O}\equiv \hat{\bm p}\cdot(\hat{\bm q}_1-\hat{\bm q}_2)\,\hat{\bm p}\cdot\frac{\hat{\bm q}_1\times \hat{\bm q}_2}{|\hat{\bm q}_1\times \hat{\bm q}_2|}.
\label{O}
\end{eqnarray}
Then we define its expectation value which is an experimental observable:
\begin{eqnarray}                       
&&\langle{\cal O}\rangle\equiv \frac{1}{N}\int \rho^{}_{M'M}{\cal O}d\Gamma^{}_{MM'}(\psi'\to J/\psi\,\pi\pi),\nonumber\\
&&N\equiv \int \rho^{}_{M'M}d\Gamma^{}_{MM'}(\psi'\to J/\psi\,\pi\pi),
\label{<O>}
\end{eqnarray}
where $\rho^{}_{M'M}$ is the density matrix, and $M$ ($M'$) stands for the magnetic quantum numbers of $\psi'$. At the $\psi'$ resonance, the energy of $e^+~(e^-)$ is $M^{}_{\psi'}/2$ wihch is much larger than the electron mass. Thus the colliding $e^+~(e^-)$ behaves essentially as a massless fermion. With the standard couplings for the process $e^+e^-\to \gamma^*\to \psi'$, a right-handed $e^+$ can only annihilate with a left-handed $e^-$ and vice versa. The resulting density matrix for $\psi'$ is
\begin{eqnarray}                             
&&\rho^{}_{M'M}=\frac{1}{2}\left(\delta^{}_{M'M}-\delta^{}_{M'0}\delta^{}_{M0}\right)\nonumber\\
&&M',M \in \{1,0,-1\}.
\label{rho}
\end{eqnarray}
See Sec.\,2.1 in ref.\,\cite{Nachtmann} for the analogous process $e^+e^-\to Z$ and set $g^{}_{Ae}=0$ there to obtain Eq.\,(\ref{rho}).
In this case the normalization constant $N$ in (\ref{<O>}) is just the total transition rate $\Gamma(\psi'\to J/\psi\,\pi\pi)$ obtained in Sec.\,IV. Since the CEDM contributions to $\Gamma(\psi'\to J/\psi\,\pi\pi)$ is negligibly small as can be seen in FIG.\,\ref{BestFit}, we can simply take the 0th order transition rate (\ref{psi'rate}) or even the experimental value $\Gamma(\psi'\to  J/\psi\,\pi\pi)=156.04\pm5.78~{\rm keV}$ [cf. Eq.\,(\ref{GammaExpt})] for the normalization constant in the following calculation.

Since ${\cal O}$ is CP odd, a nonzero $\langle{\cal O}\rangle$ can only be contributed from the CP odd part of $d\Gamma^{}_{MM'}(\psi'\to J/\psi\,\pi\pi)$, i.e., the interference terms between the E1E1 transition amplitude and the CEDM transition amplitudes. From the angular part of the phase-space integration, we can see that only the $1^3D_1\to 1^3S_1\,\pi\pi$ part in the E1E1 transition gives nonvanishing contribution to the E1E1-E1M1 interference term in (\ref{<O>}), while both the $2^3S_1\to 1^3S_1\,\pi\pi$ and $1^3D_1\to 1^3S_1\,\pi\pi$ parts in the E1E1 transition can give nonvanishing contributions to the E1E1-M1CEDM1 and E1E1-E1CEDM2 interference terms. Thus the result will take the form
\begin{widetext}
\begin{eqnarray}                         
\langle{\cal O}\rangle&=&\frac{{\cal A}}{\Gamma(\psi'\to J/\psi\,\pi\pi)}\bigg\{{\cal K}^{}_{M1CEDM1}{\cal I}^{2S\to 1S-2S\to 1S}_{E1E1-M1CEDM1}
+{\cal K}^{}_{E1CEDM2}{\cal I}^{2S\to 1S-1D\to 1S}_{E1E1-E1CEDM2}\nonumber\\
&&+\frac{{\cal C}}{{\cal A}}\bigg[{\cal K}^{}_{E1M1}{\cal I}^{1D\to 1S}_{E1E1-E1M1}
+{\cal K}^{}_{M1CEDM1}{\cal I}^{1D\to 1S-2S\to 1S}_{E1E1-M1CEDM1}
+{\cal K}^{}_{E1CEDM2}{\cal I}^{1D\to 1S-2S\to 1S}_{E1E1-E1CEDM2}\bigg]\bigg\}\delta_c,
\label{<O>form}
\end{eqnarray}
\end{widetext}
where the ${\cal I}$'s are the phase-space integrations of the interference terms which can be calculated from the approach similar to those in Sec.\,IV. So, with the measured value of
$\langle{\cal O}\rangle$, we can determine $\delta_c$ from (\ref{<O>form}). Our obtained results are:
\begin{widetext}
\begin{eqnarray}                                
&&\null\hspace{-1cm}{\cal I}^{2S\to 1S-2S\to 1S}_{E1E1-M1CEDM1}=\cos^2\theta \frac{2\delta_c}{15\pi^3m^2_c}\frac{M^{}_{J/\psi}}{\Delta M}f^{111}_{2010}f^{000}_{2010}
  \int \sin\beta(1-\cos\beta)(q_1q_2^2+q_2q_1^2)\left(q_1^{\mu}q_{2\mu}+\frac{{\cal B}}{{\cal A}}\omega_1\omega_2\right)d\omega_1d\omega_2,    \nonumber\\
&&\null\hspace{-1cm}{\cal I}^{2S\to 1S-1D\to 1S}_{E1E1-E1CEDM2}=\sin\theta \cos\theta \frac{M^{}_{J/\psi}}{ 75\sqrt{2}\pi^3}\frac{\delta_c}{m_c}f^{111}_{2010}f^{111}_{1210}
  \int \sin\beta(1-\cos\beta)(q_1q_2^2+q_1^2q_2)\left(q_1^{\mu}q_{2\mu}+\frac{{\cal B}}{{\cal A}}\omega_1\omega_2\right)d\omega_1d\omega_2,    \nonumber\\
&&{\cal I}^{1D\to 1S}_{E1E1-E1M1}=-\sqrt{\frac{3}{2}}\sin\theta \frac{M^{}_{J/\psi}S_{E1M1}}{900 \pi^3m^{}_c}
    f^{111}_{1210}\int \sin\beta(1-\cos\beta)q_1q_2(\omega_1q_2+\omega_2q_1)d\omega_1d\omega_2,       \nonumber\\
&&\null\hspace{-1cm}{\cal I}^{1D\to 1S-2S\to 1S}_{E1E1-M1CEDM1}=-\sin\theta \cos\theta\frac{\sqrt{2}M^{}_{J/\psi}}
{1575\pi^3}\frac{\delta_c}{m_c^3}f^{111}_{1210}f^{000}_{2010}\int \sin\beta(3-2\cos\beta-\cos^2\beta)q_1^2q_2^2(q_1+q_2)d\omega_1d\omega_2,    \nonumber\\
&&\null\hspace{-1cm}{\cal I}^{1D\to 1S-2S\to 1S}_{E1E1-E1CEDM2}=\sin^2\theta\frac{(\sqrt{3}-1)M^{}_{J/\psi}}{2250 \pi^3}\frac{\delta_c}{m_c}|f^{111}_{1210}|^2
  \int \sin^3\beta(q_2^3q^2_1+q_1^3q^2_2) d\omega_1d\omega_2,\nonumber\\
&&\cos\beta\equiv \frac{{\bm q}_1\cdot{\bm q}_2}{|{\bm q}_1||{\bm q}_2|}=\frac{\bigg[M^{2}_{\psi'}-M^{2}_{J/\psi}+2m^{2}_\pi-2M^{}_{\psi'}(\omega_1+\omega_2)+2\omega_1\omega_2\bigg]}
{2\sqrt{\omega_1^2-m^{2}_\pi}\sqrt{\omega_2^2-m^{2}_\pi}},
\label{Iformulas}
\end{eqnarray}
\end{widetext}
where $S_{E1M1}$ is given in (\ref{S(E1M1)}). The complicated integrations can be carried out numerically, and they lead to the following numerical results in the the CK and Cornell models:
\begin{widetext}
\begin{eqnarray}                        
&&\null\noindent{\rm CK~model}:\nonumber\\
&&~~~~\langle{\cal O}\rangle=\frac{{\cal A}}{10^3}\bigg\{2.534{\cal K}^{}_{M1CEDM1}
-0.964{\cal K}^{}_{E1CEDM2}\nonumber\\
&&~~~~~~~~~~~~+\frac{{\cal C}}{{\cal A}}\bigg[0.0123{\cal K}^{}_{E1M1}
+0.0715{\cal K}^{}_{M1CEDM1}
+0.321{\cal K}^{}_{E1CEDM2}\bigg]\bigg\}\delta_c,~~~~\nonumber\\
&&\null\noindent{\rm Cornell~model}:\nonumber\\
&&~~~~\langle{\cal O}\rangle=\frac{{\cal A}}{10^3}\bigg\{1.243{\cal K}^{}_{M1CEDM1}
-0.397{\cal K}^{}_{E1CEDM2}\nonumber\\
&&~~~~~~~~~~~~+\frac{{\cal C}}{{\cal A}}\bigg[0.00508{\cal K}^{}_{E1M1}
+0.0290{\cal K}^{}_{M1CEDM1}
+0.321{\cal K}^{}_{E1CEDM2}\bigg]\bigg\}\delta_c,~~~~
\label{<O>values}
\end{eqnarray}
\end{widetext}
in which the values of $|{\cal A}|$ and $|{\cal C}/{\cal A}|$ in the two models are given in (\ref{B/A,C/A}), and the values of ${\cal K}^{}_{E1M1}$, ${\cal K}^{}_{M1CEDM1},\,{\cal K}^{}_{E1CEDM2}$ are given in TABLE\,\ref{Kvalues} in APPENDIX\,B.

Now we come again to the problem  of the uncertain sign similar to those discussed in Sec.\,IV. We know that, in QCD, there is in principle no sign ambiguity in the various contributions in $\langle{\cal O}\rangle$ in (\ref{<O>values}). But in the present phenomenological approach to the hadronization factors, only the absolute values of the parameters, $\left|{\cal A}\right|,\,\left|{\cal C}/{\cal A}\right|,\,\left|{\cal K}^{}_{E1M1}\right|,\,\left|{\cal K}^{}_{M1CEDM1}\right|$, and $\left|{\cal K}^{}_{E1CEDM2}\right|$, can be determined as in Sec.\,III. So, in practice,  each of the five parameters has an uncertain sign. The uncertain sign of ${\cal A}$ serves as an overall uncertain sign on the right-hand-side (R.H.S.) of (\ref{<O>values}), which makes us unable to determine the sign of $\delta_c$. Despite of the overall uncertain sign, the values of the curly brackets on the R.H.S. of (\ref{<O>values}) will be affected by the uncertain signs of ${\cal C}/{\cal A},\,{\cal K}^{}_{E1M1},\,{\cal K}^{}_{M1CEDM1}$, and ${\cal K}^{}_{E1CEDM2}$. The largest term in the curly brackets in (\ref{<O>values}) is the first term. Without losing generality, we can always take ${\cal K}^{}_{M1CEDM1}>0$ with the uncertain sign of ${\cal A}$ taken into account. If we take ${\cal C}/{\cal A},{\cal K}^{}_{E1M1},{\cal K}^{}_{E1CEDM2}<0$, the curly brackets in (\ref{<O>values}) will take their largest value, 6.43 (CK model) and 3.53 (Cornell model). If we take ${\cal C}/{\cal A}<0$ while ${\cal K}^{}_{E1M1},{\cal K}^{}_{E1CEDM2}>0$, the curly brackets will take their smallest value, 5.50 (CK Model) and 3.08 (Cornell model). So the uncertainty of the values of the curly brackets caused by the uncertain signs of the SPA coefficients is $25\%$ (CK Model) and $13\%$ (Cornell model). This is better than that obtained in Sec.\,IV. Note that the uncertainties in Sec.\,IV and Sec.\,V are caused by the uncertain signs of different terms.

Moreover, we may define another related observable. Define the asymmetry based on the CP-odd operator ${\cal O}$ as
\begin{eqnarray}                          
A^{}_{{\cal O}}\equiv \frac{N_{events}({\cal O}>0)-N_{events}({\cal O}<0)}{N_{events}({\cal O}>0)+N_{events}({\cal O}<0)}.
\label{A_O}
\end{eqnarray}
This may also be used to determine $\delta_c~(d'_c)$ experimentally.

 So far there is no data on $\langle{\cal O}\rangle$. We expect BESIII to measure it.

\section{Summary and Discussions}

If the c-quark has an anomalous color-electric dipole moment (CEDM), it will serve as a new source of CP violation. In this paper, we study the
determination of size $\delta_c$ ($d'_c$) of the CEDM from the BESII data on the $M^{}_{\pi\pi}$ distribution in $\psi'\to J/\psi\,\pi\pi$ within the framework of QCD.

We have first studied the contributions of the CEDM to the hadronic transition process $\psi'\to J/\psi\,\pi\pi$, and determined the size $\delta_c$ ($d'_c$) of the CEDM by fitting the theoretical prediction to the BESII experimental data. The contributions are in two folds, namely the contribution of CEDM to the $c\bar c$ interaction potential which causes CP-even and CP-odd states mixing, and the contribution of CEDM to the vertices in the hadronic transition which affect the $M^{}_{\pi\pi}$ distribution in the transition. Both contributions lead to CP violation. Since CP violation is supposed to be small, we treat the CEDM effect as a perturbation throughout this paper. We studied these two kinds of contributions separately.

The perturbation calculations of the CEDM contribution to the $c\bar c$ potential and state mixings are given in Sec.\,II. The potential is the sum of the conventional potential $V^{}_0$ and the CEDM contribution $V^{}_1$. For $V^{}_0$, we take two extreme QCD motivated potentials, namely the CK potential and the Cornell potential to show the model dependence of the present approach. The expression for $V^{}_1$ is shown in Eq.\,(\ref{V_1'(r)}) in which only the second term contributes to state mixings. The obtained normalized state-mixing coefficients are given in Eq.\,(\ref{C}) [see also (\ref{C'})], and their numerical values are shown in TABLE\,\ref{Cvalues} in APPENDIX B.

The CEDM contribution to hadronic transition vertices is more complicated. The 0th order transitions are shown in Eqs.\,(\ref{2S-1Spipi}) and (\ref{1D-1Spipi}) with $C^{10}_{10}=C^{20}_{20}=C^{12}_{12}=1$. The $O(\delta_c^1)$ transitions are shown in Eqs.\,(\ref{1S-1Ppipi})$\--$(\ref{1D-1S E1-CEDM2}). The transition amplitudes of $O(\delta_c^1)$ transitions are proportional to $\delta_c$, and their transition rates are proportional to $\delta_c^2$. To the same order, we must also take account the transitions in (\ref{2S-1Spipi}) and (\ref{1D-1Spipi}) with the mixing coefficients of $O(\delta_c^2)$ in (\ref{Cexpansion}).

The calculation of the $M^{}_{\pi\pi}$ distribution is quite subtle. A transition amplitude contains two factors, namely the multipole gluon emission (MGE) factor and the hadronization (H) factor. For a given potential model, there is a systematic way of calculating the MGE factor \cite{KY81}\cite{Kuang06}. The calculation of the H factor is a highly nonperturbative problem in QCD. There are two approximation methods which can lead to the right order of magnitude of the transition rates \cite{KY81}\cite{Kuang06}\cite{KY90}\cite{KTY88}, namely the soft-pion approximation (SPA) and the two-gluon approximation (2GA). The SPA is a phenomenological approach which can correctly describe the angular relation between the two pions but it contains unknown constant coefficient(s) related to the hadronic matrix element in the H factor. The 2GA is a crude approximation which is easy to calculate but cannot describe the angular relation between the two pions correctly. Since we are dealing with the $M^{}_{\pi\pi}$ distribution which concerns the angular relation between the two pions, we have to take the SPA. However, to our experience, the ratios between two transition rates in SPA and 2GA are quite close to each other \cite{KY81}\cite{KTY88}. Thus we can use this approximate relation and the 2GA calculation to express the $O(\delta_c^1)$ SPA coefficients in terms of the $O(\delta_c^0)$ SPA Coefficients, cf. Eqs.\,(\ref{K_E1M1 expression}), (\ref{K_M1CEDM1 expression}), and (\ref{K_E1CEDM2 expression}) in Sec.\,III. Then we can predict the $M^{}_{\pi\pi}$ distribution by treating the CEDM contribution as perturbation. The $O(\delta_c^0)$ and $O(\delta_c^2)$ contributions to the $M^{}_{\pi\pi}$ distribution are given in Eqs.\,(\ref{dGamma/dM_pipi}), (\ref{dGamma/dM_pipi_E1M1_2S1P'}), (\ref{dGamma/dM_pipi-M1CEDM1}), (\ref{dGamma/dM_pipi_E1CEDM2_2s1s}), (\ref{dGamma/dM_pipi_E1CEDM2_1d1s}). (\ref{Gamma_E1M1-E1CEDM2}), (\ref{Gamma_M1CEDM1-E1CEDM2}), and (\ref{Gamma_E1CEDM2-E1CEDM2}).

We then made a best fit of our 0th order prediction (\ref{dGamma/dM_pipi}) to the BESII data (cf. FIG.\,\ref{BestFit}), which determines the best fit values of the SPA parameters $|{\cal A}|,\,{\cal B/A}$ and $|{\cal C/A}|$ shown in eq.\,(\ref{B/A,C/A}). Various CEDM contributions to the $M^{}_{\pi\pi}$ distribution are shown in FIG.\,\ref{CEDMs}. We see that the behaviors of the CEDM contributions are just the opposite to that of the 0th order contribution (cf. FIG.\,\ref{BestFit}). This is why the process $\psi'\to J/\psi\,\pi\pi$ can sensitively constrain $\delta_c$ ($d'_c$). Next we included the CEDM contributions to make the best fit up to the $O(\delta_c^2)$. It is shown that, with the CEDM contribution, the fit is slightly improved (with slightly smaller $\chi^2$), and the best fit values of $\left|\delta_c\right|$ and $\left|d'_c\right|$ are liste in TABLE\,\ref{delta_cvalue}. We see that the best fit value of $\left|\delta_c\right|$ ($\left|d'_c\right|$) is nonvanishing. However, considering the experimental errors, it is still consistent with zero. The $95\%$ C.L. upper bound of $\left|d'_c\right|$ is shown in Eq.\,(\ref{UpperBound}) which shows that the model dependence of the present approach is quite mild. Note that in the present approach, only the absolute values of the SPA coefficients in the CEDM contribution can be determined. So that each SPA coefficient still has an uncertain sign which may affect the result. This uncertainty is just due to the present phenomenological approach to the hadronization matrix element. We have discussed this uncertainty in Sec.\,IV, and the conclusion is that the uncertainty of the upper bound is $(34\--40)\%$ which is not so serious. Thus, taking this theoretical uncertainty into account, we conclude that {\it the $95\%$ C.L. upper bound of $\left|d'_c\right|$ in the present approach is $\left|d'_c\right|<3\times 10^{-14}$ e\,cm [cf. Eq.\,(\ref{FinalBound})] which is the first experimentally determined upper bound of the CEDM of the c quark.}

We have also proposed a second method for determining $\delta_c$ ($d'_c$) linearly by introducing a CP-odd operator ${\cal O}$ and measuring its expectation value $\langle{\cal O}\rangle$ in Sec.\,V. We have shown in Sec.\,V that this is a better way of determining $\delta_c~(d'_c)$ experimentally. So far there is no such a measurement. We suggest BESIII to do this experiment.

The state mixings caused by the CEDM of the c quark makes the transition rates $\psi'\to h_c\,\pi^0$ and $\psi'\to J/\psi\,\pi^0$ related to each other. Hence, in principle, the experimental data of these two transition rates may give another constraint on $d'_c$. However, the latest BESIII sample of $\psi'\to h_c\,\pi^0$  based on 106M of $\psi'$ events is still rather small since the branching ratio of $\psi'\to h_c\,\pi^0$ is $8\times 10^{-4}$ \cite{BESIII-h_c}, i.e., the statistical error in this transition rate is significantly larger than that in the present study. Furthermore, the transition $\psi'\to J/\psi\,\pi^0$ is dominated by E1M2 multipole gluon emissions, and the calculation  of this kind of hadronization matrix element is not so certain \cite{BES04}. Therefore the data of these two transition rates cannot provide a strong enough constraint on $d'_c$ comparable to the one obtained in the present study. So far the best experiment for determining the bound on $d'_c$ is $\psi'\to J/\psi\,\pi\pi$ at BES.


The BES detector has already been updated to BESIII with the efficiency of measuring low momentum pions significantly improved relative to BESII. So far BESIII has accumulated $1.06\times 10^8$ $\psi'$ events, and will be able to increase to $\color{DarkBlue}(7\--10)\times 10^8$ $\psi'$ events in 2012. That will be a huge sample. We expect that the new BESIII data may determine $\delta_c$ to a higher precision.

Estimating the CEDM from some UV theories may be interesting for future studies. 

\null\noindent{\bf Acknowledement}

We are grateful to Gang Li for providing us the original BESII data on $\psi'\to J/\psi\,\pi\pi$. We would like to thank Chien Yeah Seng for joining the early stage calculations in this study. This work is supported by National Natural Science Foundation under Grant Nos:10635030, 10875064, 11135003, 10975169, and 11021092.

\null\hspace{0.4cm}
\begin{center}
\bf APPENDIX A: Calculation of the Matrix Element
 $\bm{_0\langle1^1P_1|V_1|n^{(2s+1)}L_1\rangle_0}$
 \end{center}

\null\vspace{-0.4cm}
We show here the explicit expression for the relevant matrix element
$_0\langle 1^1P_1|V_1|n^{(2s+1)}(L_i)_1\rangle_0$.

In the nonrelativistic limt, the state $|n^{(2s+1)}(L_i)_1\rangle_0$ can be decomposed into the radial,
angular, and spin factors
\begin{eqnarray}                  
\Phi_{nlmm_s}(r,\theta,\phi)=R_{nl}(r)Y_l^m(\theta,\phi)X^{(2s+1)}_{m_s},
~~~~~~~~~~~~(A1)\nonumber
\label{Phi}
\end{eqnarray}
where $R_{nl}(r)$ is the radial wave function obtained from solving the Schr\"odinger equation
with a potential without $V_1$, and the spin state $X^{(2s+1)}_{M_s}$ is
\begin{eqnarray}                        
X^{(3)}_1&=&\chi_{1/2}\bar\chi_{1/2}=
\begin{pmatrix}
1\\ 0
\end{pmatrix}
\begin{pmatrix}
1\\ 0
\end{pmatrix},\nonumber\\
X^{(3)}_0&=&\frac{1}{\sqrt{2}}[\chi_{1/2}\bar\chi_{-1/2}+\chi_{-1/2}\bar\chi_{1/2}]\nonumber\\
&=&\frac{1}{\sqrt{2}}\left[
\begin{pmatrix}
1\\ 0
\end{pmatrix}
\begin{pmatrix}
0\\ 1
\end{pmatrix}
+
\begin{pmatrix}
0\\ 1
\end{pmatrix}
\begin{pmatrix}
1\\ 0
\end{pmatrix}\right],\nonumber\\
X^{(3)}_{-1}&=&\chi_{-1/2}\bar\chi_{-1/2}=
\begin{pmatrix}
0\\ 1
\end{pmatrix}
\begin{pmatrix}
0\\ 1
\end{pmatrix},\nonumber\\
X^{(0)}&=&\frac{1}{\sqrt{2}}[\chi-{1/2}\bar\chi_{-1/2}-\chi_{-1/2}\bar\chi_{1/2}]\nonumber\\
&=&\frac{1}{\sqrt{2}}\left[
\begin{pmatrix}
1\\ 0
\end{pmatrix}
\begin{pmatrix}
0\\ 1
\end{pmatrix}-
\begin{pmatrix}
0\\ 1
\end{pmatrix}
\begin{pmatrix}
1\\ 0
\end{pmatrix}\right].
\hspace{2cm}(A2)\nonumber
\end{eqnarray}

There is a spin-dependent factor $(\bm\sigma-\bar{\bm\sigma})\cdot\bm r/r$ in $V_1$. We can take the
spherical coordinate
\begin{eqnarray}                             
&&\hspace{-2cm}\frac{r_+}{r}\equiv \frac{1}{\sqrt{2}}\frac{x_1+ix_2}{r}=-\sqrt{\frac{4\pi}{3}}Y_1^1(\theta,\phi),\nonumber\\
&&\hspace{-2cm}\frac{r_-}{r}\equiv \frac{1}{\sqrt{2}}\frac{x_1-ix_2}{r}=\sqrt{\frac{4\pi}{3}}Y_1^{-1}(\theta,\phi),
\nonumber\\
&&\hspace{-2cm}\frac{r_0}{r}\equiv \frac{x_3}{r}=\sqrt{\frac{4\pi}{3}}Y_1^0(\theta,\phi),\nonumber
\end{eqnarray}
\begin{eqnarray}
&&\hspace{-2cm}\sigma_\pm\equiv \frac{1}{\sqrt{2}}(\sigma_1\pm i\sigma_2),~~~~
\bar\sigma_\pm\equiv \frac{1}{\sqrt{2}}(\bar\sigma_1\pm i\bar\sigma_2),\nonumber\\
&&\hspace{-2cm}\sigma_0\equiv \sigma_3~~~~\bar\sigma_0 \equiv\bar\sigma_3\nonumber\\
&&\hspace{-2cm}\sigma_+(\bar\sigma_+)=
\begin{pmatrix}
0~~\sqrt{2}\\ 0~~~~0
\end{pmatrix},~~~~\sigma_-(\bar\sigma_-)
=
\begin{pmatrix}
0~~~~0\\ \sqrt{2}~~0
\end{pmatrix},\nonumber\\
&&\hspace{-2cm}\sigma_0(\bar\sigma_0)=
\begin{pmatrix}
1~~~~0\\ 0~~-1
\end{pmatrix},\hspace{3.5cm}
(A3)\nonumber
\label{PolCoor}
\end{eqnarray}
and express it as
\begin{eqnarray}                              
&&\hspace{-1.5cm}(\bm{\sigma}-\bar{\bm\sigma})\cdot\frac{\bm{r}}{r}
=(\sigma_0-\bar\sigma_0)\frac{r_0}{r}+(\sigma_+-\bar\sigma_+)\frac{r_-}{r}\nonumber\\
&&\hspace{-1.5cm}~~~~+(\sigma_--\bar\sigma_-)\frac{r_+}{r}\nonumber\\
&&\hspace{-1.5cm}~~=\sqrt{\frac{4\pi}{3}}
\left\{(\sigma_0-\bar\sigma_0)Y_1^0(\theta,\phi)+(\sigma_+-\bar\sigma_+)Y_1^{-1}(\theta,\phi)
\right.\nonumber\\
&&\hspace{-1.5cm}~~\left.-(\sigma_--\bar\sigma_-)Y_1^1(\theta,\phi)\right\}.\hspace{3cm}(A4)\nonumber
\end{eqnarray}
It is easy to see that
\begin{eqnarray}                            
&&\hspace{-2cm}(\sigma-\bar\sigma)_mX^3_{m_s}=2a_m(m_s)X^1,\nonumber\\
&&\hspace{-2cm}
a_m(m_s)=(-m_s\frac{1-m_s}{2},1-|m_s|,-m_s\frac{1+m_s}{2}),\nonumber\\
&&\hspace{-2cm}\hspace{2.4cm}m=+~~~~~~m=0~~~~~~m=-\nonumber\\
&&\nonumber\\
&&\hspace{-2cm}(\sigma-\bar\sigma)_mX^1=2a^*_m(m_s)X^3_{m_s},\nonumber\\
&&\hspace{-2cm}a^*_m(m_s)=(-m_s\frac{1+m_s}{2},1-|m_s|,-m_s\frac{1-m_s}{2}),\nonumber\\
&&\hspace{-2cm}\hspace{2.2cm}m=+~~~~~~m=0~~~~~~m=-~\hspace{0cm}(A5)\nonumber
\label{sigmaX}
\end{eqnarray}
and $a_m(m_s),~a^*_m(m_s)$ satisfies
\begin{eqnarray}                           
&&\hspace{-1cm}\sum_{mm_s}a_m(m_s)a_m(m_s)=3,~\sum_{mm_s}a^*_m(m_s)a^*_m(m_s)=3,\nonumber\\
&&\hspace{-1cm}\sum_{mm_s}a^*_m(m_s)a_m(m_s)=1.\hspace{3cm}(A6)\nonumber
\label{aa}
\end{eqnarray}
Thus
\begin{eqnarray}                           
&&\hspace{-1cm}(\bm\sigma-\bar{\bm\sigma)}\cdot\frac{\bm{r}}{r}X^3_{m_s}=\sqrt{\frac{4\pi}{3}}\left\{
2a_0(m_s)Y_1^0(\theta,\phi)\right.\nonumber\\
&&\hspace{-0.8cm}\left.+2a_+(m_s)Y_1^{-1}(\theta,\phi)
-2a_-(m_s)Y_1^1(\theta,\phi)\right\}X^1\nonumber\\
&&\hspace{-1cm}(\bm\sigma-\bar{\bm\sigma)}\cdot\frac{\bm{r}}{r}X^1=\sqrt{\frac{4\pi}{3}}\left\{
2a^*_0(m_s)Y_1^0(\theta,\phi)\right.\nonumber\\
&&\hspace{-0.8cm}\left.+2a^*_+(m_s)Y_1^{-1}(\theta,\phi)
-2a^*_-(m_s)Y_1^1(\theta,\phi)\right\}X^3_{m_s}.(A7)\nonumber
\label{sigma.rX}
\end{eqnarray}
This explicitly shows that $(\bm\sigma-\bar{\bm\sigma})\cdot\bm{r}/r$ flips the quarkonium spin.

Next we evaluate the angular integration. What we need to evaluate is the integration of the product
of 3 spherical harmonics. According to the property of the spherical harmonics, we have

\begin{eqnarray}                           
&&\hspace{-0.8cm}\int~Y_{l_f}^{m_f}(\theta,\phi)
Y_1^{m}(\theta,\phi)Y_{l_i}^{-m_i}(\theta,\phi)~d\Omega\nonumber\\
&&~~=(-1)^{-m_i}\sqrt{\frac{3(2l_i+1)(2l_f+1)}{4\pi}}\nonumber\\
&&~~\times\begin{pmatrix}
l_f~1~l_i\\ 0~0~0
\end{pmatrix}
\begin{pmatrix}
l_f~~~~1~~~~~~l_i\\ m_f~~~m~-m_i
\end{pmatrix}.
\hspace{1.5cm}(A8)\nonumber
\label{YYY}
\end{eqnarray}
The values of some relevant $3-j$ symbils are
\begin{eqnarray}                            
&&\hspace{-0.5cm}
\begin{pmatrix}
1~~~~1~~~~0\\ m_f~~m~~0
\end{pmatrix}
=(-1)^{1+m}~\frac{\delta_{m_f,-m}}{\sqrt 3},\nonumber\\
&&\hspace{-0.5cm}
\begin{pmatrix}
1~~~~1~~~~2\\ 0~~~~0~~~~0
\end{pmatrix}
=\sqrt{\frac{2}{15}},\nonumber\\
&&\hspace{-0.5cm}\begin{pmatrix}
1~~~~1~~~~2\\ m_f~~m~~m_i
\end{pmatrix}=(-1)^{m_f+m}~\delta_{m_i,-m_f-m}\nonumber\\
&&\hspace{-0.5cm}~~\times\sqrt{\frac{(2+m_f+m)!(2-m_f-m)!}{30(1+m_f)!(1-m_f)!(1+m)!(1-m)!}}.\hspace{0cm}(A9)\nonumber
\label{3-j}
\end{eqnarray}

Finally we evaluate the radial integration. There are two terms in $V_1(\bm r)$ [cf. Eq.\,(\ref{V_1'(r)})]. We first look at the
first term.
\begin{eqnarray}                     
&&\hspace{-0.4cm}\frac{4}{3}\frac{\delta_c}{m^{}_c}
\int_0^\infty~\frac{g^{}_s(r)}{4\pi}R^*_{11}(r)\frac{\delta(r)}{r}R_{n_il_i}(r)~r^2dr\nonumber\\
&&~~=
\frac{8}{3}\frac{g^{}_s(r)}{4\pi}\frac{\delta_c}{m^{}_c}rR^*_{11}(r)R_{n_il_i}(r)\bigg|_{r=0}=0.
\hspace{0.4cm}(A10)\nonumber
\end{eqnarray}

\null\noindent
Here we have considered the running of the QCD coupling constant $g^{}_s(r)$ in the radial integration. Note that $g^{}_s(r)$ is governed by asymptotic freedom as $r\to 0$, i.e. $\displaystyle g^2_s(r)\stackrel{r\to 0}\sim \frac{1}{\ln(\Lambda^{}_{\overline{MS}}r)}$ (cf. Eqs.\,(A14) and (A13) below).
Eq.\,(A10) shows that the first term in Eq.\,(\ref{V_1'(r)}) actually does not contribute. 
We should only take into
account the contribution of the second term in Eq.\,(\ref{V_1'(r)}) to the matrix element. The radial integration from the second term
contribution is
\begin{eqnarray}                       
&&\hspace{-0.5cm}-\frac{4}{3}\frac{\delta_c}{m^{}_c}\int_0^\infty~\frac{g^{}_s(r)}{4\pi}R^*_{11}(r)R_{n_il_i}(r)~dr
=-\frac{4}{3}\frac{\delta^{}_c}{m^{}_c}I^{11}_{n^{}_il^{}_i}\nonumber\\
&&\hspace{-0.5cm}~~{I}_{n^{}_il^{}_i}^{11}\equiv \int_0^\infty~\frac{g^{}_s(r)}{4\pi}{R}^*_{11}(r){R}_{n^{}_il^{}_i}(r)~dr.
\hspace{2.2cm}(A11)\nonumber
\label{radial}
\end{eqnarray}
The radial wave function ${R}(r)$
is to be obtained by solving the Schr\"odinger equation.

 To include nonperturbative contributions to $g^{}_s(r)$ near the $J/\psi$ and $\psi'$ scales phenomenologically, we take the CK potential model \cite{CK} which has both a clear QCD interpretation and successful phenomenological predictions. The CK potential reads \cite{CK}
\begin{eqnarray}                      
\displaystyle
V(r)&=&-\frac{16\pi}{25}\frac{1}{rf(r)}\left[1+\frac{2\gamma_E+
\displaystyle\frac{53}{75}}{f(r)}-\frac{462}{625}\frac{\ln f(r)}{f(r)}\right]\nonumber\\
&&+kr,\hspace{5.6cm}(A12)\nonumber
\label{CKpot}
\end{eqnarray}
where $k=0.1491~{\rm GeV}^2$ is the string tension related to the Regge slope
\cite{CK}, $\gamma_E$ is the Euler constant, and $f(r)$ is
\begin{widetext}
\begin{eqnarray}                      
\displaystyle
f(r)=\ln\left[\frac{1}{\Lambda_{\overline{MS}}~r}+4.62-\bigg(1-\frac{1}{4}
\frac{\Lambda_{\overline{MS}}}{\Lambda^I_{\overline{MS}}}\bigg)
\frac{1-\exp\bigg\{-\bigg[15\bigg(3\displaystyle
\frac{\Lambda^I_{\overline{MS}}}{\Lambda_{\overline{MS}}}
-1\bigg)\Lambda_{\overline{MS}}~r\bigg]^2\bigg\}}
{\Lambda_{\overline{MS}}~r}\right]^2,~~~~~~~~~~~~(A13)\nonumber
\label{f(r)}
\end{eqnarray}
\end{widetext}
in which $\Lambda^I_{\overline{MS}}=180$ MeV. The nonperturbative effects resides in the phenomenological function $f(r)$. Writing the potential (A12)
in the standard form $V(r)=-\displaystyle\frac{4}{3}\frac{\alpha^{}_s(r)}{r}+kr$, we read from (A12) that
\bea                                   
\alpha^{}_s(r)&\equiv&\displaystyle\frac{g^2_s(r)}{4\pi}\nonumber\\
&=&\frac{12\pi}{25}\frac{1}{f(r)}\left[1+\frac{2\gamma_E+
\displaystyle\frac{53}{75}}{f(r)}-\frac{462}{625}\frac{\ln f(r)}{f(r)}\right].~(A14)\nonumber
\label{alpha_s(r)}
\eea
This running formula will be used in the calculation of $I^{11}_{n^{}_il^{}_i}$ in (A11).

Putting all the above results together, we obtain the expressions for $_0\langle 1^1P_1|V_1|n_i^3S_1\rangle_0$ and
$0\langle 1^1P_1|V_1|1^3D_1\rangle_0$:
\null\vspace{-0.2cm}
\begin{eqnarray}                            
\hspace{-0.5cm}_0\langle 1^1P_1|V_1|n_i^3S_1\rangle_0~(m_f,m_s)
&=&-\frac{8}{3\sqrt 3}\frac{\delta^{}_c}{m^{}_c}
\nonumber\\
&&\hspace{-0.2cm}\times
{I}_{n^{}_i0}^{11}\delta^{}_{m^{}_fm^{}_s},\hspace{0.5cm}(A15a)\nonumber
\label{<P|V_1|S>}
\end{eqnarray}
\null\vspace{0cm}
\begin{eqnarray}                          
&&\hspace{-0.3cm}_0\langle 1^1P_1|V_1|1^3D_1\rangle_0~(m^{}_f,m^{}_i+m^{}_s)
=\frac{8}{3}{\sqrt\frac{2}{3}}\frac{\delta^{}_c}{m^{}_c}
\nonumber\\
&&\hspace{3.2cm}~~\times {I}_{12}^{11}\delta^{}_{m^{}_f,m^{}_i+m^{}_s}.\hspace{0.2cm}(A15b)\nonumber
%
\label{<P|V_1|D>}
\end{eqnarray}
With these two matrix elements calculated, we can obtain all the mixing coefficients in
(\ref{C}) [cf. Eq.\,(\ref{C'})].

For the spin-1 states. The conventional Cartesian coordinate representation of
the spin=1 operators are:
\null\vspace{-0.2cm}
$$
S_1=\begin{pmatrix}
0~~~0~~~0\cr 0~~~0~-i\cr 0~~~i~~~0
\end{pmatrix},~
S_2=\begin{pmatrix}
0~~~0~~~i\cr 0~~~0~~~0\cr -i~0~~~0
\end{pmatrix},~
S_3=\begin{pmatrix}
0~-i~~~0\cr i~~~0~~~0\cr 0~~~0~~~0
\end{pmatrix},\eqno{(A16)}
$$
i.e.,
$$
\left(S_i\right)_{jk}=-i\epsilon_{ijk}
\eqno{(A17)}
$$
The eigenvectors of, for example, $S_3$ are:
\begin{widetext}
\begin{eqnarray}
&&\hspace{1.4cm}m_s=+1~~~~~~~~~~~~~~~~~~~~~~~~~~m_s=-1~~~~~~~~~~~~~~~~~~~~m_s=0\nonumber\\
&&\bm\chi^{}_{1m_s}=\frac{1}{2}\begin{pmatrix}
-(1+i)\cr 1-i\cr 0\end{pmatrix}~~~~~~~~
\bm\chi^{}_{1m_s}=\frac{1}{2}\begin{pmatrix}
1+i\cr 1-i\cr 0\end{pmatrix}~~~~~~~~~
\bm\chi^{}_{1m_s}=\frac{1}{2}\begin{pmatrix}
0\cr 0\cr 1\end{pmatrix},\hspace{4.95cm}(A18)\nonumber
\label{(A18)}
\end{eqnarray}
where the column is ordered according to $i=1,2,3$ from top to bottom.

In the polar coordinate system, we should make the linear combination for the component index
$(\chi^{}_{1m_s})^{}_{\pm1}=[(\chi^{}_{1m_s})^{}_1\pm i(\chi^{}_{1m_s})^{}_2]/\sqrt{2}$. Thus
we obtain the following eigenvectors in the polar coordinate system
\begin{eqnarray}
&&\hspace{1.4cm}m_s=+1~~~~~~~~~~~~~~~~~~~~~~~~~~m_s=-1~~~~~~~~~~~~~~~~~~~~m_s=0\nonumber\\
&&\bm\chi^{}_{1m_s}=\frac{1+i}{\sqrt{2}}\begin{pmatrix}
0\cr 0\cr -1
\end{pmatrix}~~~~~~~~~~~~~~
\bm\chi^{}_{1m_s}=\frac{1+i}{\sqrt{2}}\begin{pmatrix}
0\cr 1\cr 0\end{pmatrix}~~~~~~~~~
\bm\chi^{}_{1m_s}=\frac{1+i}{\sqrt{2}}\begin{pmatrix}
1\cr 0\cr 0\end{pmatrix},\hspace{3.7cm}(A19)\nonumber
\label{(A18)}
\end{eqnarray}
\end{widetext}
where the column is ordered according to $m=+1,0,-1$ from top to bottom. Compared with Eq.~(A5),
we see that
$$
\left(\chi^{}_{1m_s}\right)_m=\frac{1+i}{\sqrt{2}}a^{}_m(m_s)\equiv Na_m(m_s),
\eqno{(A19)}
$$
where $N\equiv(1+i)/\sqrt{2}$ is the normalization factor, and $N^\ast N=1$.

Now the term $\left(\chi^\ast_{1m_{s2}}\right)_j\left(S_i\right)_{jk}\left(\chi^{}_{1m_{s3}}\right)_kE_i$
in the E1CEDM2 transition amplitude can be evaluated as

\begin{eqnarray}
&&\null\hspace{-0.6cm}\left(\chi^\ast_{1m_{s2}}\right)_j\left(S_i\right)_{jk}\left(\chi^{}_{1m_{s3}}\right)_kE_i=
-i\epsilon_{ijk}\left(\chi^\ast_{1m_{s2}}\right)_j\left(\chi^{}_{1m_{s3}}\right)_kE_i\nonumber\\
&&\,\,~
=-i\left(\bm\chi^\ast_{1m_{s2}}
\times\bm\chi^{}_{1m_{s3}}\right)\cdot\bm E\nonumber\\
&&~~
=\epsilon_{-m_1,-m_2,-m_3}\left(\chi^\ast_{1m_{s2}}\right)_{m_2}
\left(\chi^{}_{1m_{s3}}\right)_{m_3}E_{m_1}\nonumber\\
&&~~
\stackrel{(A19)}=\epsilon_{-m_1,-m_2,-m_3}a^{}_{m_2}(m_{s2})a_{m_3}(m_{s3})E_{m_1}\nonumber\\
&&~~
=-i\left(\bm a^{}(m_{s2})
\times\bm a(m_{s3})\right)\cdot\bm E.\hspace{2.2cm}(A20)\nonumber
\label{(A20)}
\end{eqnarray}

\null\vspace{0.6cm}
\begin{center}
\bf APPENDIX B: Numerical Results of $\bm{f^{LP^{}_IP^{}_F}_{n^{}_Il^{}_In^{}_Fl^{}_F}}$, $\bm{C^{n_Fl_F}_{n_Il_I}}$, , and $\bm{{\cal K}}$'s
 \end{center}

Here we list the numerical values of the the mxing coefficients $C^{n_Fl_F}_{n_Il_I}$, reduced amplitudes $f^{LP^{}_IP^{}_F}_{n^{}_Il^{}_In^{}_Fl^{}_F}$,  and the SPA hadronization factor coefficients ${\cal K}$'s in the CK and Cornell models.

\null\vspace{-0.8cm}
\begin{table}[h]
\caption{Values of the mixing coefficients $C^{n_Fl_F}_{n_Il_I}$ in the CK and Cornell models.}
\tabcolsep 16pt
\begin{tabular}{ccc}
\hline\hline
&CK model&Cornell model\\
\hline
$C^{11}_{10}$&0.1165$\delta_c$&0.1142$\delta_c$\\
$C^{11}_{20}$&-0.07119$\delta_c$&-0.07209$\delta_c$\\
$C^{11}_{12}$&0.1597$\delta_c$&0.1669$\delta_c$\\
$C^{10}_{20}$&-0.01863$\delta_c$&-0.01824$\delta_c$\\
$C^{10}_{12}$&-0.03906$\delta_c$&-0.04091$\delta_c$\\
\hline\hline
\label{Cvalues}
\end{tabular}
\end{table}

\newpage
\begin{table}[h]
\label{fvalues}
\caption{Values of the reduced amplitudes $f^{LP^{}_IP^{}_F}_{n^{}_Il^{}_In^{}_Fl^{}_F}$ in the CK and Cornell models.}
\tabcolsep 12pt
\begin{tabular}{ccc}
\hline\hline
&CK model&Cornell model\\
\hline
$f^{111}_{2010}$\,(GeV$^{-3}$)&8.8715&6.9330\\
$f^{000}_{2010}$\,(GeV$^{-1}$)&0.3869&0.3709\\
$f^{111}_{1210}$\,(GeV$^{-3}$)&-11.2507&-8.7701\\
$f^{101}_{1110}$\,(GeV$^{-2}$)&-3.31446&-2.9040\\
$f^{010}_{1110}$\,(GeV$^{-2}$)&-6.8977&-6.1708\\
$f^{110}_{2011}$\,(GeV$^{-2}$)&4.8399&4.2801\\
$f^{001}_{2011}$\,(GeV$^{-2}$)&4.5643&4.0379\\
$f^{110}_{1211}$\,(GeV$^{-2}$)&-5.8859&-5.1934\\
$f^{201}_{1211}$\,(GeV$^{-2}$)&-4.1087&-3.6324\\
\hline\hline
\label{fvalues}
\end{tabular}
\end{table}
\begin{table}[h]
\caption{Values of the SPA hadronization factor coefficients $\left|{\cal K}^{}_{E1M1}\right|$. $\left|{\cal K}^{}_{M1CEDM1}\right|$, and $\left|{\cal K}^{}_{E1CEDM2}\right|$ obtained from the best fit values of $|{\cal A}|$, ${\cal B/A}$, and ${\cal C/A}$ in the CK and Cornell models.}
\tabcolsep 16pt
\begin{tabular}{ccc}
\hline\hline
&CK model&Cornell model\\
\hline
$\left|{\cal K}^{}_{E1M1}\right|$&3.289&4.208\\
$\left|{\cal K}^{}_{M1CEDM1}\right|$&2.318&2.625\\
$\left|{\cal K}^{}_{E1CEDM2}\right|$&0.532&0.602\\
\hline\hline
\label{Kvalues}
\end{tabular}
\end{table}
\end{text}

\newpage


\end{document}